# Primordial $N_2$ provides a cosmochemical explanation for the existence of Sputnik Planitia, Pluto

Christopher R. Glein [*], J. Hunter Waite Jr.

Space Science and Engineering Division, Southwest Research Institute, San Antonio, TX, United States

**Contact information**

Christopher Glein
Space Science and Engineering Division
Southwest Research Institute
P.O. Drawer 28510
San Antonio, TX 78228-0510
United States
Phone: 210-522-5510
Fax: 210-522-3547
Email: cglein@swri.edu

**Highlights**

- Now is the time to begin data-driven investigations into the origin of Pluto's $N_2$.
- We estimate the amount of $N_2$ observed and potentially lost though time.
- Models of primordial $N_2$ with an initial cometary or solar abundance of $N_2$ to $H_2O$ can satisfy the mass balance constraints.
- Consistency with the low atmospheric mixing ratio of CO may require removal of primordial CO, which can be attributed to the burial of CO in surface ices, or its destruction from exposure to liquid water.

**Abstract**

The presence of $N_2$ in the surface environment of Pluto is critical in creating Pluto's richness of features and processes. Here, we propose that the nitrogen atoms in the $N_2$ observed on Pluto were accreted in that chemical form during the formation of Pluto. We use *New Horizons* data and models to estimate the amounts of $N_2$ in the following exterior reservoirs: atmosphere, escape, photochemistry, and surface. The total exterior inventory is deduced to be dominated by a glacial sheet of $N_2$-rich ices at Sputnik Planitia, or by atmospheric escape if past rates of escape were much faster than at present. Pluto's atmosphere is a negligible reservoir of $N_2$, and photochemical destruction of $N_2$ may also be of little consequence. Estimates are made of the amount of $N_2$ accreted by Pluto based on cometary and solar compositions. It is found that the cometary model can account for the amount of $N_2$ in Sputnik Planitia, while the solar model can provide a large initial inventory of $N_2$ that would make prodigious atmospheric escape possible. These consistencies can be considered preliminary evidence in support of a primordial origin of Pluto's $N_2$. However, both models predict accreted ratios of $CO/N_2$ that are much higher than that in Pluto's atmosphere. Possible processes to explain "missing CO" that are given quantitative support here are fractional crystallization from the atmosphere resulting in CO burial at the surface, and aqueous destruction reactions of CO subject to metastable thermodynamic equilibrium in the subsurface. The plausibility of primordial $N_2$ as the primary source of Pluto's nitrogen (vs. $NH_3$ or organic N) can be tested more rigorously using future constraints on the $^{14}N/^{15}N$ ratio in $N_2$ and the $^{36}Ar/N_2$ ratio.

**1. Introduction**

With a similar general role as water on Earth, methane on Titan, and especially $CO_2$ on Mars, molecular nitrogen ($N_2$) is the key volatile on Pluto that brings activity to the frigid surface environment of this remote world (see Olkin et al., 2017 and Stern et al., 2018 for recent reviews on the Pluto system). A volatile is defined here as a chemical species that can readily transition in a macroscopic sense between gaseous and condensed forms at the temperature of a planetary body. Solid $N_2$ appears to be the most abundant ice on the surface of Pluto accessible to spectroscopy (Owen et al., 1993; Cruikshank et al., 2015; Grundy et al., 2016a; Protopapa et al., 2017). The *New Horizons* mission discovered what is inferred to be an $N_2$-rich ice sheet in a near-equatorial region called Sputnik Planitia (formerly referred to as Sputnik Planum; Stern et al., 2015), which constitutes the western lobe of Pluto's "heart" (Tombaugh Regio). Because of its relatively low viscosity at Pluto surface temperatures (~40 K), solid $N_2$ is able to deform and flow, which maintains the youthful appearance of Sputnik Planitia (McKinnon et al., 2016; Trowbridge et al., 2016), and leads to the erosion of bedrock and the formation of glacial landforms (Moore et al., 2016; Howard et al., 2017). The relatively low triple point temperature of $N_2$ (63 K) facilitates melting, which could occur at the base of glaciers (Howard et al., 2017), or in the ambient surface environment if there has been modest warming (a few tens of kelvins) via large impacts locally or global climate change (Stern et al., 2017). Because $N_2$ ice has a relatively high vapor pressure, it can readily sublimate at Pluto's surface, which initiates a volatile cycle that results in pitting of such deposits (Moore et al., 2017), seasonally dependent mass and heat transport (Earle et al., 2017), and the deposition of bright frosts (Buratti et al., 2017). This cycle is also largely responsible for the existence of an atmosphere on Pluto (Hubbard et al., 1988; Elliot et al., 1989; Yelle & Lunine, 1989; Gladstone et al., 2016; Young et al., 2018).

What was the original molecular carrier of the nitrogen atoms that are now contained in $N_2$ on Pluto? Mandt et al. (2016) studied the isotopic evolution of Pluto's nitrogen for primordial $N_2$ and $NH_3$ sources of nitrogen. It has been suggested that organic N (i.e., organic molecules containing nitrogen atoms in their chemical structures) could also be a significant carrier of N to Pluto and other Kuiper belt objects (McKinnon et al., 1997; 2008). A primordial (accretional) origin of Pluto's $N_2$ would be simple in the sense that only outgassing is required to explain the observations of $N_2$ (e.g., Owen, 1982). In contrast, an $NH_3$ source of $N_2$ requires both chemistry and outgassing, in either order depending on whether the generation of $N_2$ takes place in the interior (e.g., Glein et al., 2009), or near the surface (e.g., Atreya et al., 1978; McKay et al., 1988; Sekine et al., 2011). An organic source requires a more specific process of thermally driving the formation of $N_2$ in a putative rocky core, followed by outgassing (e.g., Miller et al., 2017). Of course, a mixed source of $N_2$ is not to be excluded. These hypotheses, borrowed from the Titan literature, view the $N_2$ as being derived from the bulk planetary inventory. A different approach is to consider exogenous mechanisms of bringing $N_2$ to the surface of Pluto (e.g., cometary impacts; Singer & Stern, 2015). Beyond an intrinsic interest in the source of $N_2$, insights into the origin and evolution of Pluto and the solar system as a whole can be elucidated by addressing the issue of the origin of $N_2$ on Pluto (Lunine, 1993a). These pertain to the composition of the building blocks of Pluto and the conditions of their formation (e.g., temperature-pressure); the thermal history of the interior, surface, and atmosphere of Pluto; and the processes responsible for similarities and differences between Pluto and other $N_2$-bearing bodies such as Titan, Triton, and Eris (e.g., Broadfoot et al., 1989; Niemann et al., 2010; Tegler et al., 2012).

The purpose of this paper is to examine the hypothesis of a primordial origin of $N_2$, which represents an effort to take a step forward in determining the origin of Pluto's $N_2$. How consistent with the available data is the notion that $N_2$ was obtained from the formation environment of Pluto in that chemical form? Owen et al. (1993) discussed their discovery of $N_2$ on Pluto's surface in the context of a primordial origin, because Pluto seems to have formed far from the Sun where temperatures are presumed to have been low enough to accrete $N_2$ (e.g., <30 K). However, such a low-temperature origin cannot be guaranteed, as Pluto could have formed inside its present orbit and been scattered outward during a period of giant planet migration (e.g., Levison et al., 2008). The relatively high bulk density of Pluto (1854 kg/m$^3$; Nimmo et al., 2017) suggests that Pluto formed beyond the giant planets (McKinnon & Mueller, 1988), but this does not necessarily mean that temperatures in its formation environment were sufficiently cold to accrete $N_2$. Stern et al. (1997) attempted to use the $N_2$/CO ratio of Pluto's surface as a cosmochemical constraint, but the high value seemed to be most indicative of volatile removal processes (e.g., atmospheric escape or hydrothermal geochemistry), which led to ambiguity between potential $N_2$ and $NH_3$ sources of the observed nitrogen.

Our understanding of the origin of Pluto's $N_2$ progressed little over the last twenty some years. However, the success of the *New Horizons* mission has changed the situation, and detailed observational data (e.g., Stern et al., 2015; Gladstone et al., 2016; Grundy et al., 2016a; Moore et al., 2016) can now be brought to bear on this problem (Section 2). It is timely to examine the primordial $N_2$ hypothesis in particular, given the recent first detection of $N_2$ from a comet by the *Rosetta* spacecraft (Rubin et al., 2015). These two datasets allow a mass balance test of this hypothesis to be performed (Section 3), based on a cosmochemical model of Pluto as a "giant comet" (for a pioneering application of this type of comparative approach to Triton, see Lunine, 1993b; Lunine et al., 1995; and McKinnon et al., 1995). We also consider the possibility that the building blocks of Pluto could have been as rich as the solar composition in terms of the abundance of $N_2$. We discuss in Section 4 physical and chemical mechanisms that might reconcile possible primordial values of the $CO/N_2$ ratio with observations of this ratio on Pluto. Lastly, we conclude this paper with a summary of our findings, and some open questions and suggestions for future observations (Section 5).

## 2. The apparent inventory of $N_2$ on Pluto

The goal of this section is to estimate the amount of $N_2$ that may have been outgassed from Pluto's interior, which can be regarded as an apparent inventory (we call it "apparent" because it is an inventory that can be quantified on the basis of our current understanding of Pluto, which is undoubtedly incomplete). Broadly, we can define two reservoirs of volatiles on Pluto: exterior and interior. We focus on Pluto's exterior because there is a lack of data that can be used to probe the $N_2$ content of its interior (e.g., Glein, 2015), where $N_2$ could be stored in clathrate hydrates, dissolved in a liquid water ocean (Hammond et al., 2016; Johnson et al., 2016; Keane et al., 2016; Nimmo et al., 2016), or trapped in a rocky core. The possibility of a global, kilometer-scale crustal layer of $N_2$ ice residing above a water ice mantle is implausible, as its existence may prevent the detection of widespread water ice on Pluto's surface (Grundy et al., 2016a), and an $N_2$ crust would be too weak to support the mountainous terrains observed by *New Horizons* (Stern et al., 2015). The exterior can be divided into sub-reservoirs that presently contain volatiles or are irreversible sinks of volatiles. We term these sub-reservoirs: atmosphere, escape, photochemistry, and surface.

*2.1. Atmosphere*

The mass of Pluto's atmosphere ($m_{atm}$) can be estimated from the atmospheric pressure at the surface ($P_{atm}$) using the following equation

$$P_{atm} \approx \frac{m_{atm} g}{4\pi R_{avg}^2}, \qquad (1)$$

where $g = 0.616$ m/s$^2$ designates the gravitational acceleration at the surface (Stern et al., 2015), and $R_{avg} = 1188$ km the average radius of Pluto (Nimmo et al., 2017). For $P_{atm} \approx 12$ μbar (~1.2 Pa; Hinson et al., 2017), the mass of the atmosphere is ~3.5×10$^{13}$ kg. This is consistent with earlier estimates that were made using Earth-based observations (e.g., ~3×10$^{13}$ kg; Singer & Stern, 2015). The calculated mass can be assumed to be essentially identical to the mass of $N_2$ in Pluto's atmosphere as the near-surface atmosphere is >99% $N_2$ by volume (Young et al., 2018). Hence, there are ~1×10$^{15}$ moles of atmospheric $N_2$ (Table 1).

**Table 1.** Simple estimates for the exterior inventories of $N_2$ on Pluto from *New Horizons* observations and models.

| Model: | Past Like Present (PLP) | Large Loss (LL) |
|---|---|---|
| Reservoir | Moles of $N_2$ | Moles of $N_2$ |
| Atmosphere | $1 \times 10^{15}$ | $1 \times 10^{15}$ |
| Escape | $5 \times 10^{16}$ | $(0.1\text{-}1) \times 10^{22}$ |
| Photochemistry | $2 \times 10^{18}$ | $5 \times 10^{18}$ |
| Surface [a] | $(0.4\text{-}3) \times 10^{20}$ | $(0.4\text{-}3) \times 10^{20}$ |
| Sum ≈ Outgassed amount | $(0.4\text{-}3) \times 10^{20}$ | $(0.1\text{-}1) \times 10^{22}$ |

[a] The surface inventory is assumed to be dominated by volatile ices in Sputnik Planitia (see Section 2.3).

*2.2. Escape and photochemistry*

It has been inferred that Pluto's atmosphere is escaping by the Jeans mechanism, with an escape rate for $N_2$ of ~$5 \times 10^{22}$ molecules/s (~$3 \times 10^6$ mol/yr; Young et al., 2018). This is several orders of magnitude slower than pre-*New Horizons* predictions (Tian & Toon, 2005; Tucker et al., 2012; Zhu et al., 2014), but follows from the cooler temperature profile that is required to fit the solar occultation data from *New Horizons* (Gladstone et al., 2016). That profile is also consistent with CO data from the ALMA (Atacama Large Millimeter/submillimeter Array) telescope (Lellouch et al., 2017). The inference of slow atmospheric escape from Pluto is supported by the finding of only a small region of interaction between Pluto's atmosphere and the solar wind (Bagenal et al., 2016). SWAP (Solar Wind Around Pluto) data also indicate that the heavy ion tail behind Pluto is most likely dominated by $CH_4^+$ rather than $N_2^+$ (Zirnstein et al., 2016), consistent with a higher escape rate for $CH_4$ (~$6 \times 10^{25}$ molecules/s; Young et al., 2018). The Young et al. (2018) value for the rate of escape of $N_2$ can be considered the current "best" estimate. Understanding escape at Pluto is a work in progress; if novel observational data (e.g., Lisse et al., 2017) or modeling of neutral and ion escape leads to a revision, then that should be incorporated in subsequent studies.

Photochemical reactions occur in Pluto's atmosphere. $C_2$ hydrocarbons and hydrogen cyanide have been detected in the atmosphere, and these are the primary photochemical products of $CH_4$ and $N_2$-$CH_4$, respectively (Gladstone et al., 2016; Lellouch et al., 2017). We obtain an estimate for the rate of destruction of $N_2$ from the photochemical model of Wong et al. (2017), which was semi-parameterized to be consistent with data from *New Horizons* (e.g., altitude profiles of species densities). This model predicts that the most abundant photochemical products derived from $N_2$ are HCN, $CH_3C_2CN$ (methylcyanoacetylene), and $HC_3N$ (cyanoacetylene), with surface precipitation fluxes of 35, 6, and 4 g cm$^{-2}$ Gyr$^{-1}$, respectively (Wong et al., 2017). The downward flux of HCN from the model has a similar value as an estimate of 24 g cm$^{-2}$ Gyr$^{-1}$ derived from the observed column density of HCN from ALMA (Lellouch et al., 2017). By accounting for the nitrogen content of the reported small molecules (Wong et al., 2017), we can recast the preceding fluxes in terms of grams of nitrogen, which can be summed to estimate the photochemical loss flux of nitrogen as 20.5 g cm$^{-2}$ Gyr$^{-1}$. This corresponds to a present $N_2$ destruction rate of ~$1 \times 10^8$ mol/yr.

To quantify the escape and photochemical inventories of $N_2$, information is needed on the rates of escape and photochemistry in the past. With regards to the present situation, cooling in the upper atmosphere dictates the rate of escape for $N_2$ (Gladstone et al., 2016; Young et al., 2018). Zhang et al. (2017) argued that tholin haze particles dominate the atmospheric radiative balance, whereas Strobel & Zhu (2017) hypothesized that the "unknown" cooling agent could be exogenic $H_2O$. To account for the effects of escape and photochemistry on the $N_2$ budget, we would need to know their average rates over the history of Pluto. The lack of knowledge of the history of Pluto's atmosphere represents a key limitation on the ability to integrate the loss of $N_2$. Nevertheless, we can consider two cases that allow an initial exploration of the possible implications of different loss regimes on the $N_2$ budget of Pluto. In a first case, we adopt the geological principle of "the present is the key to the past", and extrapolate the present rates of loss into the past by scaling them by the time-dependent solar fluxes of extreme ultraviolet and Lyman-alpha photons (see Appendix). We term this the PLP (Past Like Present) model. It is found that the average rate may be a factor of ~5 larger than the present rate. This simplistic approach implies that the average rates of escape and photochemistry in the PLP model are $\sim 1\times 10^7$ and $\sim 5\times 10^8$ mol/yr, respectively. Applying these rates over 4.56 Gyr, we find that $\sim 5\times 10^{16}$ moles of $N_2$ might have been lost to space, and $\sim 2\times 10^{18}$ moles of $N_2$ might have been converted to photochemical products (Table 1).

To obtain a much different point of reference, we specify a case with faster rates of escape and photochemistry. This is called the LL (Large Loss) model. For the rate of escape for $N_2$, we consider values between $10^{27}$ and $10^{28}$ molecules/s ($[0.5-5]\times 10^{11}$ mol/yr). This is the solar energy-limited regime, and the considered range is what was generally expected prior to the *New Horizons* flyby (Tian & Toon, 2005; Tucker et al., 2012; Zhu et al., 2014). These values are 4 to 5 orders of magnitude larger than the post-flyby estimate from Young et al. (2018). The range adopted for the LL model can be interpreted in two ways. First, it may be possible that the predictions were correct in the sense that they are the "normal" rates, while the cooler upper atmosphere observed by *New Horizons* is atypical (e.g., if the cooling mechanism is strongly dependent on season, the solar cycle, or irregular delivery of outgassed or exogenous material). Alternatively, Pluto's atmosphere may have transitioned to the present cold upper atmosphere regime relatively recently, but the atmosphere could have been in the energy-limited regime for most of Pluto's history. The LL model helps to illustrate how large rates of past escape could impact the calculated budget of $N_2$ on Pluto.

A large estimate for the amount of $N_2$ destroyed by photochemistry can be made by considering Titan as an analogue of a past Pluto that may have had a thicker atmosphere (Stern et al., 2017). We use results from the Wong et al. (2015) Titan model to facilitate a comparison to their Pluto model. The production fluxes of nitriles (dominated by HCN) predicted for present-day Titan can be summed to calculate a destruction flux for $N_2$ of $3.4\times 10^8$ molecules cm$^{-2}$ s$^{-1}$. As an approximation, we estimate the loss flux of $N_2$ if there was previously a Titan-like atmosphere at Pluto's semi-major axis by scaling the Titan loss flux by the ratio of solar fluxes; i.e., Pluto ≈ Titan × $(9.6/39)^2 = 2.1\times 10^7$ molecules $N_2$ cm$^{-2}$ s$^{-1}$. This would imply an $N_2$ destruction rate of $\sim 2\times 10^8$ mol/yr. This rate is only a factor of ~2 higher than the present rate on Pluto from Wong et al. (2017), demonstrating the dominant role of the photon flux in the photochemical destruction of $N_2$ (rather than atmospheric density). The "present" loss rates in the LL model can be multiplied by a factor of 5 to approximate the long-term rates (see Appendix). If we apply these rates over 4.56 Gyr, we obtain escape and photochemical inventories of $\sim(0.1-1)\times 10^{22}$ and $\sim 5\times 10^{18}$ moles of $N_2$, respectively (Table 1).

The PLP and LL models of $N_2$ loss are intended to provide starting points for exploring the roles of escape and photochemistry in the $N_2$ budget of Pluto. They are not endmembers, so the real range of uncertainty may be larger than the range of values in Table 1. For example, there could have been substantial escape during Pluto's early history as a result of impacts (Marounina et al., 2015; Robbins et al., 2017), including a Charon-forming giant impact (McKinnon, 1989; Canup, 2005; Sekine et al., 2017). Alternatively, the integrated loss of $N_2$ could be lower than calculated if the $N_2$ atmosphere freezes-out for a significant period of time near aphelion (e.g., Bertrand & Forget, 2016); or if $N_2$ has existed at the surface for much less than the age of the solar system (e.g., geologically recent outgassing of $N_2$ from the interior). A future measurement of the ratio of $^{36}Ar/^{38}Ar$ below the homopause could allow the amount of past atmospheric escape to be constrained, as escape processes lead to enrichment in the heavy isotope. Also, it is intriguing to ponder if the color of Charon's polar caps (Grundy et al., 2016b) can be interpreted as a record of the $N_2/CH_4$ ratio of escaped gases in Pluto's past. Insights into the history of photochemistry on Pluto could be gained if the compositions (e.g., nitriles vs. hydrocarbons) and quantities of organic solids on Pluto's surface, such as those suspected at Cthulhu Regio (Grundy et al., 2016a; Protopapa et al., 2017), can be constrained via laboratory studies and *New Horizons* observations.

*2.3. Surface/Sputnik Planitia*

We consider the surface inventory of $N_2$ on Pluto to be dominated by geological deposits (e.g., landforms containing $N_2$ throughout at least a decameter scale), while frosts (i.e., seasonal coatings on less volatile materials) may be negligible contributors on a total mass basis. Estimating the amount of $N_2$ in surface deposits requires information on their lateral and vertical extents. Spectroscopy provides clues to the lateral extent on the uppermost light-reflecting layer of the surface, but it is usually difficult to determine the thickness of a deposit remotely. A remarkable exception may be Sputnik Planitia (Stern et al., 2015; White et al., 2017), which is a crater-free, highly reflective, oval-shaped unit that is interpreted to consist of a sheet of $N_2$-rich ices that is undergoing solid-state convection (Moore et al., 2016; McKinnon et al., 2016; Trowbridge et al., 2016; Buhler & Ingersoll, 2018). The volume of ices in Sputnik Planitia can be estimated using the surface area (~870,000 km$^2$; Moore et al., 2016) and an average thickness. McKinnon et al. (2016) suggested a thickness of ~3-6 km if the ice is convecting in the sluggish lid regime, while Trowbridge et al. (2016) argued that convection occurs in the Rayleigh-Bénard regime with an ice thickness of ~10 km. These model-dependent values are determined by the diameters of polygonal cells in Sputnik Planitia (~30 km in the center of the deposit), which have different relationships to the depth of the ice depending on the style of convection (McKinnon et al., 2016; Trowbridge et al., 2016).

The range in volume implied by the thickness estimates is $(2.6-8.7)\times10^6$ km$^3$. However, this may be an upper limit for the volume of $N_2$ because (1) radiative transfer modeling suggests that $CH_4$ may be as abundant as $N_2$ in some parts of Sputnik Planitia (Protopapa et al., 2017), (2) the ice sheet is thought to be thinner near its edges because the convection cells are narrower there (McKinnon et al., 2016; Trowbridge et al., 2016), and (3) the presence of CO at the surface of Sputnik Planitia (Grundy et al., 2016a) makes it possible for there to be additional CO that could be concentrated with depth (McKinnon et al., 2016; Trowbridge et al., 2016). Therefore, we also consider the possibility that the volume of $N_2$ ice may be half the previous range, which leads to a more conservative range for the volume of $N_2$ ice of $(1.3-8.7)\times10^6$ km$^3$. This estimate could be improved if the three-dimensional physical and chemical structure of the ice sheet can be constrained by geologic and compositional mapping, and convection modeling. Using the latter range in volume, an $N_2$ ice density of ~0.96 g/cm$^3$ (Moore et al., 2016), and a molar mass of 28 g/mol, we estimate an abundance of ~(0.4-3)×10$^{20}$ moles of $N_2$. If Sputnik Planitia has a bulk cometary or solar (i.e., a high) CO/$N_2$ ratio but is heavily stratified with respect to this ratio (see Section 4.1), then the abundance of $N_2$ in the deposit could be a factor of ~0.006 to ~0.1 smaller than the previous range.

It is assumed that the total inventory of $N_2$ on Pluto's surface is similar to that in Sputnik Planitia. The strongest $N_2$ infrared absorptions are observed in Sputnik Planitia (Grundy et al., 2016a; Protopapa et al., 2017). While $N_2$ ice has been detected elsewhere (Grundy et al., 2016a; Protopapa et al., 2017), it is unclear whether those ices are present as frosts or volumetrically significant geological deposits. On the other hand, Sputnik Planitia appears to act as a cold trap for mobile species like $N_2$, given that it is located at latitudes that receive minimal solar irradiance over long timescales (Hamilton et al., 2016), and it is in a deep basin where higher atmospheric pressure promotes condensation (Bertrand & Forget, 2016). Because there are mechanisms for $N_2$ to accumulate in Sputnik Planitia, this feature may be the dominant reservoir of $N_2$ ice on Pluto. Studies of compositional and geologic features (e.g., topography) outside of Sputnik Planitia will be needed to assess if there are other deposits that could be significant to the global surface inventory of $N_2$.

Our estimates for the apparent inventory of $N_2$ on Pluto are given in Table 1. The atmosphere is a negligible reservoir of $N_2$. In the PLP model, escape and photochemistry are minor factors, while the surface deposit at Sputnik Planitia contains the largest quantity of $N_2$ (equivalent to an atmospheric pressure of ~0.4-3 bar). In the LL model, escape is the dominant reservoir of $N_2$ (the amount lost corresponds to an atmospheric pressure of ~10-100 bar), followed by Sputnik Planitia and photochemistry. The total exterior inventory in the LL model is ~3-250 times larger than that in the PLP model (Table 1). The full range of these models is (0.4-100)×10$^{20}$ moles of $N_2$ ([0.1-30]×10$^{19}$ kg).

**3. Cosmochemical models for the accreted inventory of $N_2$**

In this section, we attempt to constrain the amount of $N_2$ that could have been present in the building blocks of Pluto. To do this, we first estimate the amount of $H_2O$ on Pluto from the bulk density of the body. We then consider a compositional model based on cometary observations, as well as more $N_2$-rich models to help understand what the primordial (molar) ratio of $N_2$/$H_2O$ might have been for Pluto. The accreted inventory of $N_2$ from these approaches can be compared to the outgassed inventories from the PLP and LL models (Table 1) to assess if primordial sources of $N_2$ can account for the latter inventories.

*3.1. The $H_2O$ content of Pluto*

A two-component model of water (liquid+ice) and rock is used to represent the bulk composition of Pluto (cf., McKinnon & Mueller, 1988). For a mechanical mixture, the volumetric mass density of the mixture ($\rho_{\text{avg}}$) can be expressed in terms of the constituent densities as

$$\rho_{\text{avg}}^{-1} = f_w \rho_w^{-1} + f_r \rho_r^{-1}, \tag{2}$$

where $f_i$ stands for the mass fraction of constituent i (w = water, r = rock), and $\rho_i$ its density. By way of mass balance ($f_w + f_r = 1$), Equation 2 can be rearranged to

$$f_w = \frac{(\rho_r / \rho_{avg}) - 1}{(\rho_r / \rho_w) - 1}, \tag{3}$$

which allows the mass fraction of water inside Pluto to be estimated. To evaluate Equation 3, we adopt the following density values: $\rho_{avg}$ = 1854 kg/m$^3$ (Nimmo et al., 2017), $\rho_w$ = 920-1000 kg/m$^3$, and $\rho_r$ = 3050-3530 kg/m$^3$. The range in $\rho_w$ spans an interior that is dominated by ice Ih to one that may contain a deep ocean of liquid water (Hammond et al., 2016; Johnson et al., 2016; Keane et al., 2016; Nimmo et al., 2016). Assuming that Pluto is differentiated into a rocky core overlain by a water layer of liquid+ice, Hammond et al. (2016) suggested that the core should be denser than ~3050 kg/m$^3$, because radiogenic heating would make the inner volume of the core too hot to permit the existence of a larger proportion of lower-density hydrated silicates. We adopt the mean density of Io as an upper limit for the density of completely dehydrated chondritic rock (i.e., silicate+metal; Schubert et al., 2007).

Using the above density values, we obtain a water fraction of 0.28 to 0.36 of the total mass of Pluto. This is consistent with a recent estimate of 0.34 by McKinnon et al. (2017a). Our approach does not account for water that has been lost from the hydrosphere by reactions with rock (e.g., serpentinization; Glein et al., 2015), nor the possible decrease in rock density if it contains a significant organic component (which would lead to a decrease in $f_w$; McKinnon et al., 1997). An intermediate value for the water fraction of 0.32 can be used to approximate the global abundance of water, as uncertainty in the amount of $N_2$ accreted by Pluto is dominated by uncertainty in the primordial ratio of $N_2/H_2O$ (see Sections 3.2 and 3.3). For a Pluto mass of 1.3×10$^{22}$ kg (Stern et al., 2015) and a nominal $H_2O$ fraction of 0.32, we estimate an $H_2O$ abundance of 4.2×10$^{21}$ kg (2.3×10$^{23}$ moles).

*3.2. Cometary constraints on the primordial $N_2/H_2O$ ratio*

We set potential limits on the initial $N_2/H_2O$ ratio of Pluto, based on the supposition that the compositions of known comets provide clues to the bulk composition of the icy planetesimals that accreted into Pluto (e.g., Lunine & Nolan, 1992). As an example, McKinnon et al. (2017b) pointed out that the grain density of comet 67P/Churyumov-Gerasimenko (~1800-1900 kg/m$^3$) is similar to the uncompressed density of Pluto (~1820 kg/m$^3$). The mixing ratio of $N_2$ has been measured in the coma of only one comet (67P; Rubin et al., 2015), so we do not have definitive data on how variable the $N_2$ abundance may be among comets. However, significant variability can be expected given the incredible diversity of comets in terms of the abundances of other chemical species (e.g., Mumma & Charnley, 2011; Le Roy et al., 2015). There is unlikely to be a single "cometary abundance" of $N_2$, but it may be possible to specify a plausible range that can be applied to Pluto.

The mixing ratio of $N_2$ in cometary ices can be expressed as ($N_2/H_2O$) = ($N_2/CO$)×($CO/H_2O$). It is assumed that comets may have similar $N_2/CO$ ratios, and the value for comet 67P ([5.70±0.66]×10$^{-3}$; Rubin et al., 2015) is representative. This is a necessary assumption because there is only one measurement. In an attempt to be more inclusive, we make a provisional assumption that this ratio may deviate from the 67P range by a factor of two among comets. In situ measurements of $N_2$ in other comets are desirable. If future measurements were to show a wider range of variability in the $N_2$ abundance among comets, then the uncertainty range for our cometary model (see below) would need to be widened proportionally. We adopt a range for the $CO/H_2O$ ratio of 1.4-8.8%, which corresponds to the 99% confidence interval for 24 comets from Dello Russo et al. (2016), who term this range "typical values". This statistically significant range may be appropriate for the initial inventory of Pluto, because a body as big as Pluto would form from a large number of comet-like icy planetesimals. A small population of atypical composition may not contribute appreciably to the bulk composition.

We adopt the typical range of $CO/H_2O$ for all comets analyzed by Dello Russo et al. (2016), and do not make any assumptions about whether the value for any particular comet or cometary family should be most similar to the composition of Pluto. The CO mixing ratio in the coma of comet 67P from *Rosetta* spans a wide range with values of 2.7% and 20% above the summer and winter hemispheres of the comet, respectively (Le Roy et al., 2015). We suspect that the former value is more representative of the bulk abundance of CO in the nucleus, because there should be less fractionation between CO and $H_2O$ due to their difference in volatility at higher temperatures in the summer. Thus, comet 67P is likely to fall in the adopted range (see above).

We multiply the number of moles of $H_2O$ inside Pluto (see Section 3.1) by the $N_2/H_2O$ ratio from the above cometary model to estimate the accreted inventory of $N_2$ (Figure 1). Based on the previous cometary range, it is predicted that Pluto could have started with $\sim(0.8\text{-}26)\times10^{19}$ moles of $N_2$. This is generally consistent with the apparent inventory from the PLP model, which suggests that an inventory of $N_2$ brought to Pluto by comet-like icy planetesimals is sufficient to explain the existence of $N_2$ in Sputnik Planitia (Table 1). It is remarkable that such a simple approach can get within even an order of magnitude of what is needed by the PLP model. There is only a small region of the comet-based parameter space (cyan region in Figure 1) where the $N_2$ supply would not be large enough to overlap the PLP range for the apparent inventory. In the magenta region (Figure 1), there is agreement between the amount of $N_2$ available and the amount required by the PLP model. If Sputnik Planitia has a bulk cometary $CO/N_2$ ratio but most of the CO is buried (see Section 4.1), then the lower limit on the PLP inventory would be $\sim2\times10^{17}$ moles of $N_2$. This would decrease the demand on the theoretical model such that the entire comet-based parameter space would provide more $N_2$ than the minimum abundance in the PLP inventory.

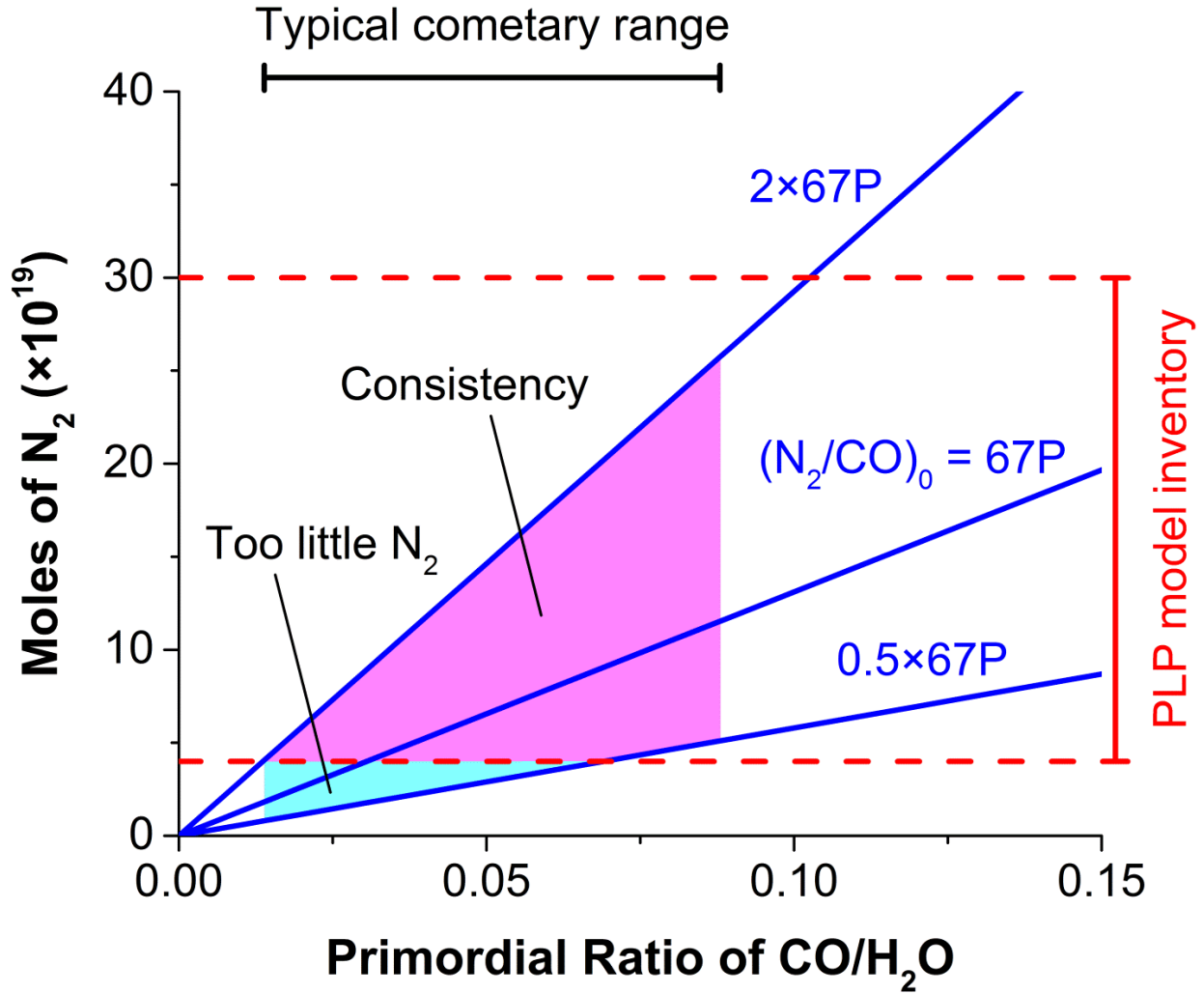

**Figure 1.** Cometary model for the inventory of primordial $N_2$ on Pluto (blue lines) as a function of the $N_2/CO$ and $CO/H_2O$ ratios of accreted ices. The primordial ratio of $N_2/CO$ is expressed relative to that in comet 67P/Churyumov-Gerasimenko (Rubin et al., 2015), which is assumed to be representative of comet-like icy planetesimals to within a factor of two. The typical range of cometary CO is from Dello Russo et al. (2016), and this range is assumed to be representative for the building blocks of Pluto. The area delineated by the dashed red lines designates the PLP model inventory (Table 1). The whole shaded region represents the theoretically permitted parameter space for the cometary model. This region is divided into cyan and magenta sub-regions to indicate how the amount of $N_2$ available in these sub-regions compares to the exterior inventory of $N_2$ in the PLP model. This type of direct comparison implicitly assumes that all of the accreted $N_2$ would be outgassed.

The cometary model does not provide sufficient primordial $N_2$ to achieve consistency with the LL model (Figure 2). If the LL model approximates Pluto's past more closely than does the PLP model, this discrepancy would imply that a source providing more primordial $N_2$ than the cometary model must be invoked, or the source of $N_2$ on Pluto was a different nitrogen-bearing species that is more abundant in comets (e.g., Mandt et al., 2014; Miller et al., 2017).

The cometary model could not provide enough accreted $N_2$ to account for the surface inventory (Table 1) if more than ~$22\times10^{19}$ moles of $N_2$ have escaped (mean rate >$9\times10^{26}$ molecules/s over 4.56 Gyr). Because the cometary model cannot maintain consistency with the observational constraint for a large amount of escape, there should be minimal escape-induced isotopic fractionation if this is an appropriate model for Pluto. As an example, the $^{38}Ar/^{36}Ar$ ratio on Pluto should be similar to the average cometary value in this model. This ratio has only been measured in the coma of comet 67P where $^{38}Ar/^{36}Ar \approx 0.19$ (Balsiger et al., 2015). In the rest of this discussion, we focus on the cometary model in terms of its consistency with the PLP inventory (we return to the LL model in Section 3.3).

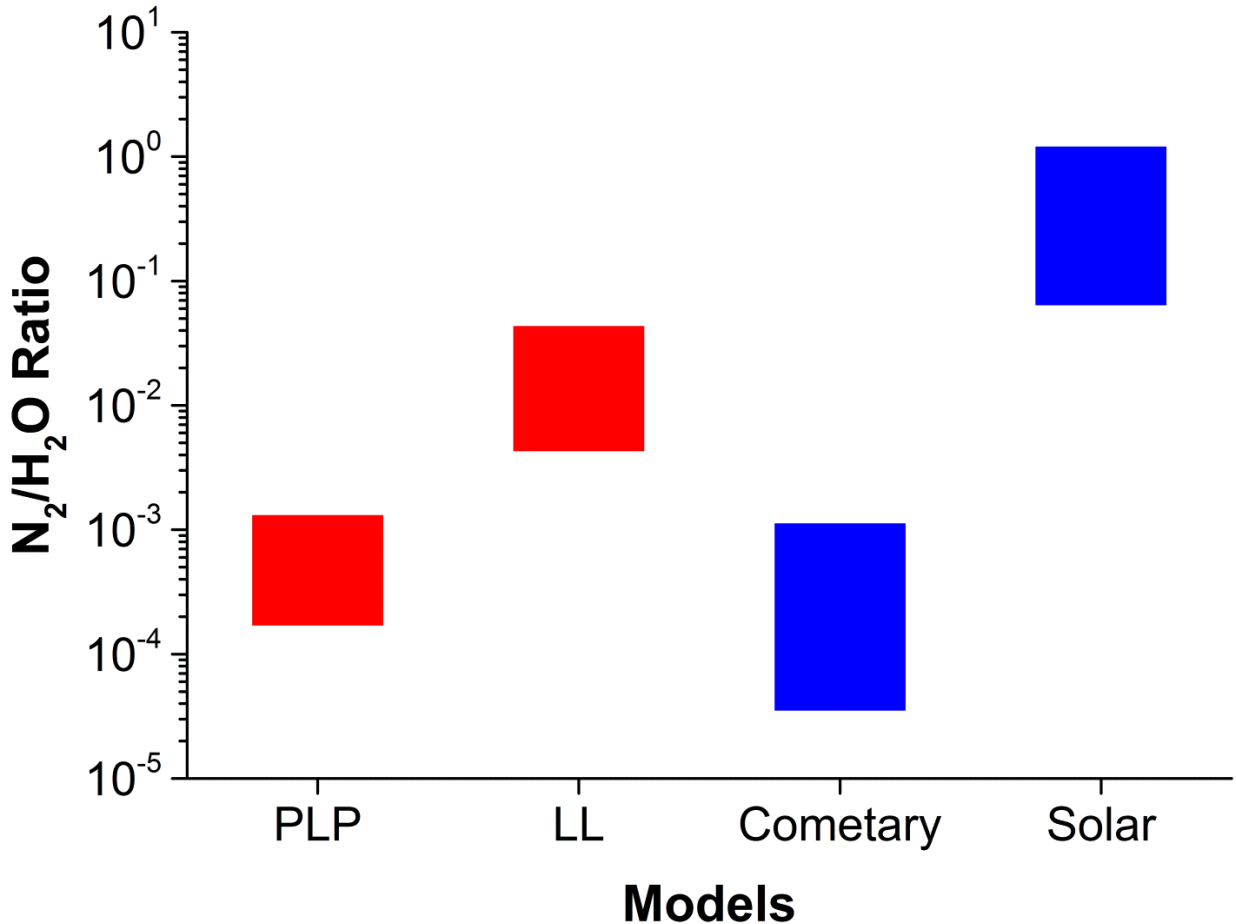

**Figure 2.** A comparison of top-down (red) and bottom-up (blue) model estimates for the ratio of $N_2/H_2O$ on Pluto. The "top-down" class of models is rooted in data from Pluto with an assumption on the history of $N_2$ loss (see Section 2.2), while the "bottom-up" class represents a theoretical construction of Pluto from a type of primordial building block (see Sections 3.2 and 3.3). Note that neither the PLP nor the LL model attempt quantification of possible subsurface $N_2$.

For an upper limit of 100% outgassing of primordial $N_2$, the cometary model suggests that the surface inventory on Pluto may be no larger than ~$26\times10^{19}$ moles. A surface inventory smaller than this could be explained if the accreted ratio of $N_2/H_2O$ was lower than the cometary maximum in Figure 2 ($\sim 10^{-3}$), atmospheric escape was a few orders of magnitude faster in the past, or if the outgassing of $N_2$ is incomplete. Provided that the cometary model is appropriate for Pluto, the outgassing efficiency ($\varepsilon$ = amount outgassed/total amount) may not be arbitrarily low. Using the minimum exterior inventory ($\sim 4 \times 10^{19}$ moles) and the maximum cometary inventory ($\sim 26 \times 10^{19}$ moles), we can set a potential lower limit on the outgassing efficiency for $N_2$ of $\sim 15\%$. However, the outgassing efficiency could be lower if some of the observed $N_2$ is not primordial (i.e., derived from $NH_3$ or organic N). Information on the outgassing efficiency would be useful for deciphering the geophysical evolution of Pluto, such as its potential cryovolcanic history (Neveu et al., 2015; Moore et al., 2016; Singer et al., 2016). On the other hand, if geophysical arguments were to give preference to an outgassing efficiency significantly lower than $\sim 15\%$, then the cometary model cannot provide sufficient primordial $N_2$ to account for the inferred surface inventory (Table 1).

We also consider several examples of special interest. First, if it is assumed that Pluto's initial $N_2$/CO ratio was the same as in comet 67P, the primordial ratio of $CO/H_2O$ would need to be higher than $\sim 3.1\%$ to be consistent with the lower limit on the PLP exterior inventory of $N_2$. In a second case, we find that Pluto's initial $N_2$/CO ratio would need to be greater than $\sim 0.6$ times the mean value in comet 67P, if the proto-Plutonian ices had $CO/H_2O$ ratios similar to the average cometary value (i.e., 5.2%; Dello Russo et al., 2016). Third, if we assume a case with a 67P $N_2$/CO ratio and an average cometary $CO/H_2O$ ratio, we can predict the amount of $N_2$ accreted by Pluto, which would be $\sim 7 \times 10^{19}$ moles. This could be regarded as a nominal estimate that would be at the lower end of the PLP range for outgassing in Table 1, but nonetheless consistent with it. In this case, the minimum outgassing efficiency would be increased to $\sim 60\%$, perhaps suggestive of intense cryovolcanism. This would be supported if a future in situ mission were to find that a relatively large fraction of Pluto's radiogenic $^{40}$Ar or $^{129}$Xe has been outgassed (e.g., Tobie et al., 2012; Glein, 2017). In general, we find that the cometary model is able to account for the exterior inventory of $N_2$ on Pluto, provided that the outgassing efficiency is high and the escaped inventory is small.

*3.3. Possibilities for a higher primordial ratio of $N_2/H_2O$*

The accreted inventory of $N_2$ on Pluto could have been larger than predicted by our empirical cometary model (see Section 3.2). One way for this to happen is if we underestimated the accreted ratio of $CO/H_2O$ in that model. It has been suggested that nonpolar volatile species in comets were accreted as clathrate hydrates (e.g., Mousis et al., 2016). If so, a possible theoretical upper limit on the $CO/H_2O$ ratio in icy planetesimals would correspond to complete occupancy of clathrates by CO (i.e., $CO \cdot 5.75H_2O$), which would yield a $CO/H_2O$ ratio of $\sim 0.17$. This would have the effect of increasing the upper limit from the cometary model by a factor of $\sim 2$, which would allow up to $\sim 51 \times 10^{19}$ moles of $N_2$ to be accreted by Pluto. If Pluto accreted ices that were completely clathrated, sufficient $N_2$ would have been brought in to account for the largest PLP inventory (Table 1). However, not even the smallest LL inventory can be accommodated by a scenario of complete clathration (Table 1). Overall, this scenario would provide a slight upward extension to the previous cometary model, but the conclusion of consistency with PLP and inconsistency with LL is maintained.

The initial inventory of primordial $N_2$ on Pluto could have been huge if Pluto accreted $N_2$-rich amorphous ices or $N_2$ ice itself. In these cases, an upper limit on the amount of $N_2$ accreted can be determined from solar system elemental abundances (Owen & Encrenaz, 2006).

The strongest evidence supporting a model of solar $N_2$ is the discovery of a nitrogen-rich extrasolar object (proposed to be a Kuiper belt analogue) that is being accreted by a white dwarf star (Xu et al., 2017). $N_2$-rich comets have yet to be found. Upper limits on the ratios of $N_2/H_2O$ and $CO/H_2O$ in a solar model can be estimated by assuming that all N and C in the solar nebula were present as $N_2$ and CO, consistent with the inhibited kinetics of reducing these species to $NH_3$ and $CH_4$, respectively (Lewis & Prinn, 1980). Using the solar abundances of Palme et al. (2014), we compute an $N_2/H_2O$ ratio ranging from ~0.13 to ~1.2 (with $N_2/CO \approx 0.08$-0.17). By scaling a nominal $N_2/H_2O$ ratio of ~0.3 by Pluto's $H_2O$ abundance (see Section 3.1), we calculate a corresponding inventory of ~$7\times10^{22}$ moles (equivalent to a ~130 km thick global surface layer of $N_2$ ice). For a lower limit, we adopt an $N_2/H_2O$ ratio of ~0.064 (i.e., half of the minimum solar abundance), consistent with the dominance of $N_2$ in the solar nebular nitrogen budget (Owen et al., 2001), while recognizing the possibility of other significant carriers of N in the outer solar system (i.e., $NH_3$, organic materials).

The solar $N_2$ model provides such a large amount of primordial $N_2$ that it does not seem consistent with the PLP inventory (Figure 2). Achieving consistency between these models would require the outgassing efficiency of $N_2$ to be very low ($\varepsilon < 2\%$), such that almost all of the primordial $N_2$ would have to remain hidden in Pluto's interior. It is unclear if minimal outgassing of $N_2$ can be consistent with the suspected occurrence of cryovolcanism on Pluto (Neveu et al., 2015; Moore et al., 2016; Singer et al., 2016). Conversely, the solar model provides an $N_2$ inventory that is a closer match to the LL inventory (Figure 2). In this comparison, $\varepsilon$ can be as large as ~70%, which may be more plausible for outgassing of a volatile primordial species over the history of Pluto.

Storing excess $N_2$ in Pluto's interior provides one way of reconciling the solar and LL models. The difference could also be explained if the escape rate in the LL model is an underestimate for the average rate over 4.56 Gyr, as the actual amount of escaped $N_2$ is unknown (see Section 2.2). An upper limit on the average rate of escape can be obtained by assuming 100% outgassing, and dividing the solar inventory by $4.56\times10^9$ yr. This results in a nominal rate of order $10^{13}$ moles/yr (~$10^{29}$ molecules/s). This approach of treating the average escape rate as a free parameter ($r_{esc}$) can be generalized using

$$\varepsilon n_0 \approx n_{surf} + r_{esc} t , \qquad (4)$$

which assumes that the major reservoirs of outgassed $N_2$ are the surface and escape (see Table 1); and where $n_0$ designates the initial inventory (e.g., ~$7\times10^{22}$ moles), $n_{surf}$ the present surface inventory, and $t$ the duration of escape (i.e., $4.56\times10^9$ yr). Figure 3 provides a map of the combinations of escape and outgassing that would allow the solar model to be consistent with the surface inventory at Sputnik Planitia.

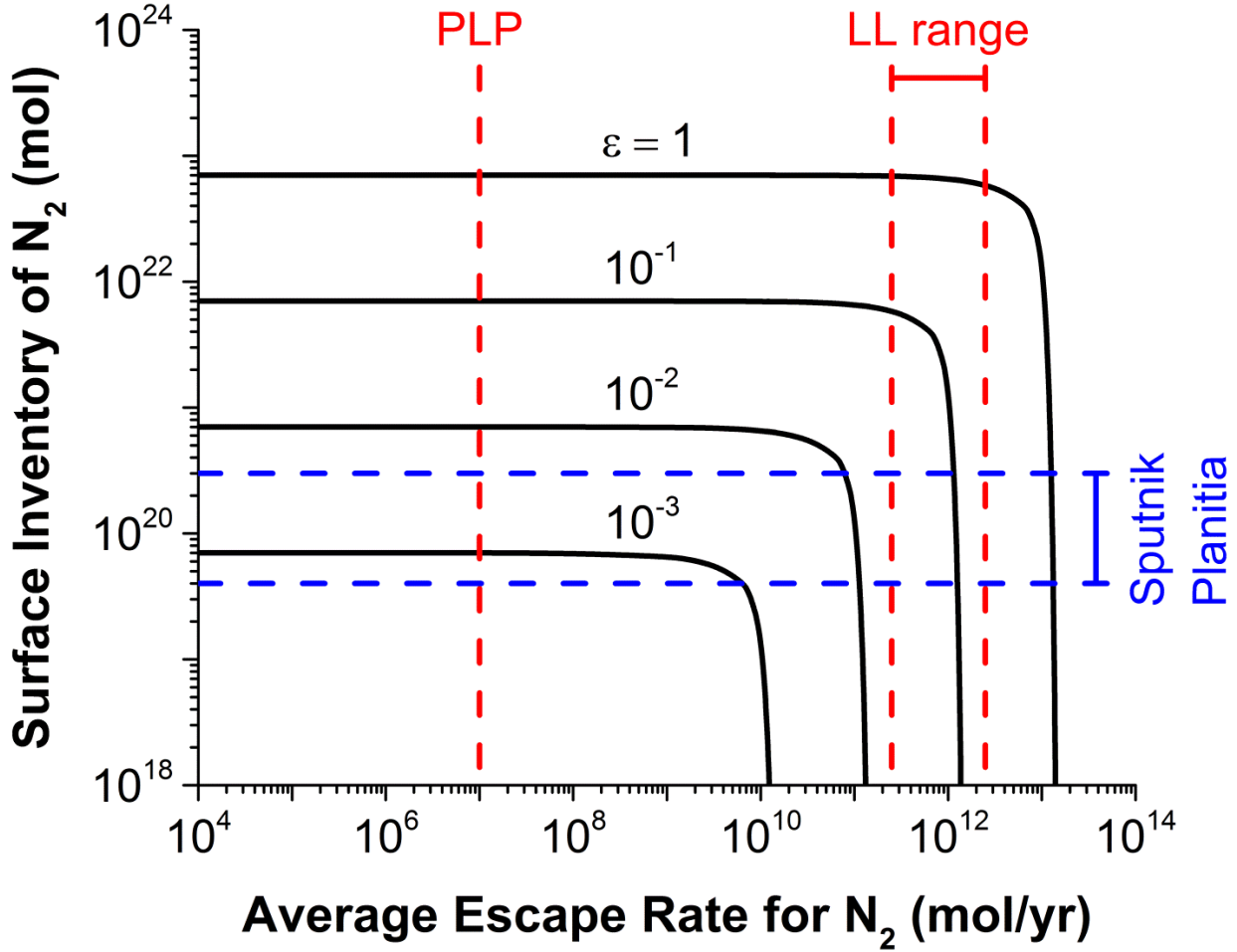

**Figure 3.** Relationships between Pluto's surface inventory of $N_2$, its escape rate, and outgassing efficiency (contour values) for a model with an accreted solar ratio of $N_2/H_2O$ (here, the initial amount is $7 \times 10^{22}$ mol $N_2$). The dashed red lines show rates for the PLP and LL models (see Section 2.2), and the dashed blue lines bracket the estimated inventory of $N_2$ in Sputnik Planitia (Table 1). The average rate is over 4.56 Gyr.

A third way to obtain a match to the LL inventory is to treat the accreted $N_2/H_2O$ ratio for Pluto as a free parameter, or as a mixture between the cometary and solar models. We do not elaborate on such scenarios because they are difficult to test due to the lack of specificity. It is more productive to focus on well-defined cases, such as the cometary and solar models. However, this possibility underscores the importance of the accreted inventory to the mass balance, and its implications for outgassing and escape.

We conclude that the solar model can explain the origin of $N_2$ on Pluto, but it provides too much $N_2$ unless the outgassing efficiency is low or past rates of escape were rapid. Independent data are needed to assess if these requirements are realistic. This model may suggest a lack of radiogenic $^{40}Ar$ and $^{129}Xe$ on Pluto's surface or in the atmosphere, indicative of limited outgassing. Prodigious atmospheric escape could be revealed by a significantly super-solar ratio of $^{38}Ar/^{36}Ar$ (solar = 0.182; Pepin et al., 2012). These measurements must await a future Pluto lander equipped with a mass spectrometer. In the meantime, modeling of escape from possible past atmospheres, as well as heat and mass transfer out of different internal structures would help to constrain Pluto's history of escape and outgassing.

## 4. A case of missing CO

As discussed in Sections 3.2 and 3.3, the cometary model can account for the amount of $N_2$ in the PLP inventory, while the solar model shows closer agreement with the LL inventory. These initial consistencies suggest that there is potential for accretion of $N_2$ to be responsible for the origin of $N_2$ on Pluto, with respect to mass balance. However, a long-standing and unresolved question relating to a primordial origin of Pluto's $N_2$ is: Where is all the CO that would be accreted with $N_2$ (Owen et al., 1993)? In the cometary model, the $CO/N_2$ ratio is ~175 (see Section 3.2), and $CO/N_2 \approx 9$ in the solar model (see Section 3.3). In contrast, the ratio of $CO/N_2$ in Pluto's atmosphere is ~$5 \times 10^{-4}$ (Lellouch et al., 2017). Both models of primordial $N_2$ have $CO/N_2$ ratios that are too high by several orders of magnitude. The low ratio in the atmosphere of Pluto could be attributed to: (1) physical enrichment of primordial $N_2$; (2) chemical depletion of primordial CO; (3) escape of primordial $N_2$ and CO, followed by production of $N_2$ from a less volatile precursor (Lunine & Nolan, 1992); or (4) non-accretion of $N_2$ and CO, also followed by $N_2$ production (Mandt et al., 2014; Miller et al., 2017). Below, we investigate the feasibility of the first two hypotheses.

*4.1. Sequestering CO in surface ices*

In the burial hypothesis, the discrepancy in the $CO/N_2$ ratio could be explained if ices on the surface of Pluto are substantially enriched in CO compared with the atmosphere, or if the outgassing efficiency of CO from the interior is much less than that for $N_2$. We focus on the first possibility because a large amount of fractionation is required, and fractionation resulting from differences in volatility is larger at lower temperatures (Pluto's surface is the coldest part of the body). Below, we consider two endmember models of physical fractionation.

For a model of complete (batch) equilibrium between solid and gas, we can calculate the atmospheric ratio of $CO/N_2$ from the initial value using

$$(CO/N_2)_{atm} = \frac{(CO/N_2)_0}{(1-F_{N_2,atm})K + F_{N_2,atm}}, \qquad (5)$$

which accounts for surface + atmosphere mass balance and solid-gas equilibrium of CO and $N_2$, and $F_{N2,atm}$ represents the fraction of $N_2$ remaining in the atmosphere. The equilibrium constant for $CO(g) + N_2(s) = CO(s) + N_2(g)$ is given by

$$K = \frac{p^o_{N_2}}{p^o_{CO}} \approx \frac{(CO/N_2)_{solid}}{(CO/N_2)_{gas}}, \qquad (6)$$

in which Raoult's law is assumed to be obeyed, and $p_i^o$ refers to the vapor pressure of pure solid i. The assumption that $N_2$ and CO exhibit ideal solid solution behavior is supported by the CO-$N_2$ binary excess Gibbs energy data of Lobo & Staveley (1985), which indicate that ices in equilibrium with Pluto's atmosphere should have a CO/$N_2$ ratio consistent with the Raoult's law value to within a factor of two. A representative value of $K \approx 8$ can be calculated by adopting a surface temperature of ~37 K near Sputnik Planitia (Hinson et al., 2017). At this temperature, the vapor pressures of pure $N_2$ and CO ices are ~12 µbar and ~1.5 µbar, respectively (Fray & Schmitt, 2009).

Figure 4 shows the evolution of the CO/$N_2$ ratio in Pluto's atmosphere from the model of batch equilibrium. It can be seen that this fractionation process is not efficient enough to decrease the CO/$N_2$ ratio sufficiently to yield the observed value from an initial cometary or solar ratio. For the limiting case of $F_{N2,atm} \to 0$, the atmospheric CO/$N_2$ ratio from batch equilibrium is only ~8 (or $K$) times smaller than the starting value, not several orders of magnitude smaller as observed at Pluto (Lellouch et al., 2017). The Sputnik Planitia glacier would have a CO/$N_2$ ratio (~$4\times10^{-3}$) that is much lower than the cometary and solar values, if all of the ice is at equilibrium with the atmosphere (the ice may be slightly more depleted in CO due to non-ideal behavior). The surface concentration of CO at Sputnik Planitia has not yet been quantified using *New Horizons* data (Grundy et al., 2016a), but the Raoult's law value is consistent with a previous model of the global near-infrared spectrum (CO/$N_2 \approx 5\times10^{-3}$; Owen et al., 1993). On the other hand, Sputnik Planitia could be greatly enriched in CO at depth (McKinnon et al., 2016; Trowbridge et al., 2016) if it is layered by volatility, as CO has a significantly lower vapor pressure than $N_2$ on Pluto (see above).

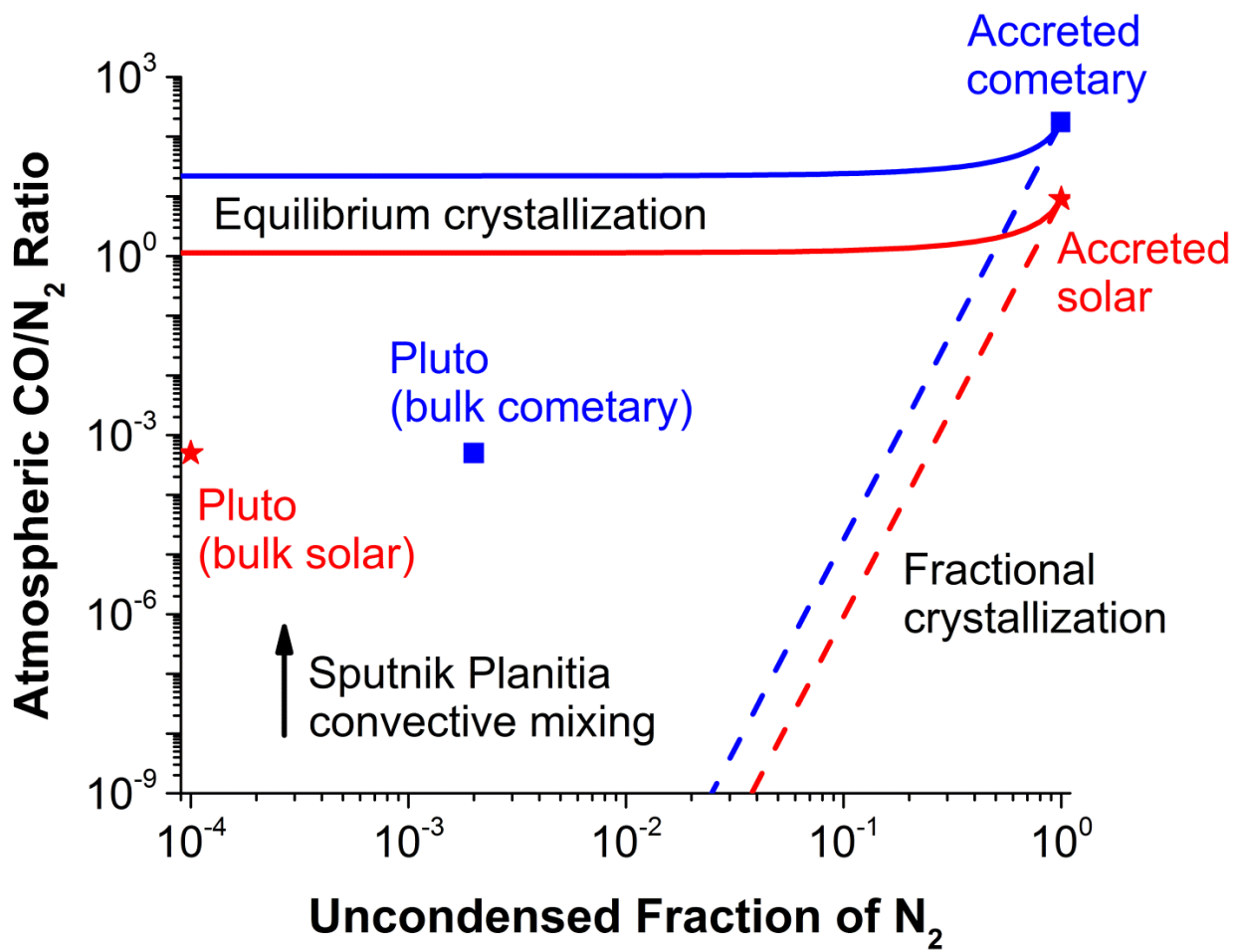

**Figure 4.** Endmember models of CO/N$_2$ fractionation by gas to solid condensation at 37 K on Pluto. Starting from a cometary (upper right blue square) or solar (upper right red star) CO/N$_2$ ratio, the lines depict evolutionary trajectories for the atmospheric CO/N$_2$ ratio as a function of the fraction of N$_2$ remaining in the atmosphere. The solid lines are for equilibrium crystallization of these atmospheric gases, while the dashed lines show predictions for fractional crystallization. The symbols labeled "Pluto" are based on the present atmospheric ratio of CO/N$_2$ (Lellouch et al., 2017), and on estimates of the uncondensed fraction of N$_2$ if Sputnik Planitia has a bulk cometary or solar ratio of CO/N$_2$ (see Section 4.1). One way to understand the positions of these points falling between the endmember lines is if the depth profile of ices in Sputnik Planitia is determined by a balance between layering based on volatility and mixing by solid-state convection (e.g., Figure 5).

As another endmember, volatile layering can be modeled as a fractional crystallization process using one of the Rayleigh equations

$$(CO/N_2)_{atm} = (F_{N_2,atm})^{K-1} \times (CO/N_2)_0. \tag{7}$$

This approach assumes that previously condensed material can no longer equilibrate with the atmosphere as a result of burial. In Figure 4, we find that Rayleigh fractionation of a cometary or solar CO/N$_2$ ratio overlaps Pluto's atmospheric CO/N$_2$ ratio for $F_{N2,atm}$ values of ~0.16 or ~0.25, respectively. However, these values are inconsistent with the estimated amount of volatile ices in Sputnik Planitia (~10$^{20}$ moles of N$_2$+CO; see Section 2.3). For an atmospheric N$_2$ abundance of ~1×10$^{15}$ moles (Table 1) and assuming that the glacier has a bulk cometary or solar CO/N$_2$ ratio, we estimate $F_{N2,atm}$ values of ~2×10$^{-3}$ and ~1×10$^{-4}$ for the cometary and solar models, respectively. These values result in far too much fractionation of the CO/N$_2$ ratio (Figure 4). Atmospheric CO would be undetectable for this model.

Equilibrium crystallization of CO and N$_2$ from the atmosphere is not efficient enough at fractionating the CO/N$_2$ ratio, while fractional crystallization is too efficient. Does this mean that fractionation by condensation cannot be invoked to explain why the CO/N$_2$ ratio in Pluto's atmosphere is far below the cometary and solar values? The answer is no because an intermediate scenario could lead to some volatile layering that is partially counteracted by convective mixing in Sputnik Planitia (McKinnon et al., 2016; Trowbridge et al., 2016). Because the Pluto symbols in Figure 4 fall between the endmember curves, a hybrid scenario could allow the cometary and solar models to achieve consistency with the present set of constraints.

As an exercise in seeing how large of an effect convective mixing might need to make, we can treat the quantity $K$ in Equation 7 as an adjustable parameter. We find that the values for the cometary and solar models would need to be 3.05 and 2.06, respectively, to reproduce the atmospheric ratio of CO/N$_2$ using the $F_{N2,atm}$ values given above. Because these values are significantly smaller than the thermodynamic equilibrium value of ~8 (see Equation 6) and they occur as exponents in Equation 7, convection is suggested to play an important role that would be equivalent to decreasing the vapor pressure ratio of N$_2$/CO by a factor of ~3-4. Treating $K$ in this way can be useful, as this approach provides some initial quantitative insight by allowing this type of comparison to be made. However, it must be cautioned that Equation 7 becomes an assumed fitting function in this approach, which can no longer be directly traced to the physics of fractional crystallization.

It is of interest to illustrate the possible implications of such a model of modest Rayleigh-type fractionation for the compositional profile of ices in Sputnik Planitia. The CO/N$_2$ ratio in a given layer of these ices can be computed as

$$(\text{CO/N}_2)_{\text{ice}} = K \times \left(F_{\text{N}_2,\text{atm}}\right)^{K-1} \times (\text{CO/N}_2)_0 . \tag{8}$$

A convenient way to relate $F_{\text{N2,atm}}$ to the height ($z$) of the layer of interest (as measured from the bottom of the ice sheet) is as follows

$$\frac{z}{H} = \frac{1 - F_{\text{N}_2,\text{atm}}(z)}{1 - F_{\text{N}_2,\text{atm}}(H)} , \tag{9}$$

where $H$ stands for the thickness of the ice sheet. By combining Equations 8 and 9, we can calculate the CO/N$_2$ ratio in the ice as a function of the dimensionless height (Figure 5). The ice at the bottom of the deposit is calculated to be modestly enriched in CO (by a factor of $K$), whereas the uppermost layers would be greatly enriched in N$_2$ (by several orders of magnitude). While based on empirical fitting of $K$, the profiles in Figure 5 can be regarded as two (of many) possibilities for the subsurface of Sputnik Planitia, which are chiefly meant to call attention to large vertical gradients in composition as a key issue. The major takeaway is that a Rayleigh-type model can reconcile primordial CO/N$_2$ ratios with that in Pluto's atmosphere, because very large fractions of these volatiles are frozen out on the surface, which amplifies the decrease in the residual atmospheric CO/N$_2$ ratio.

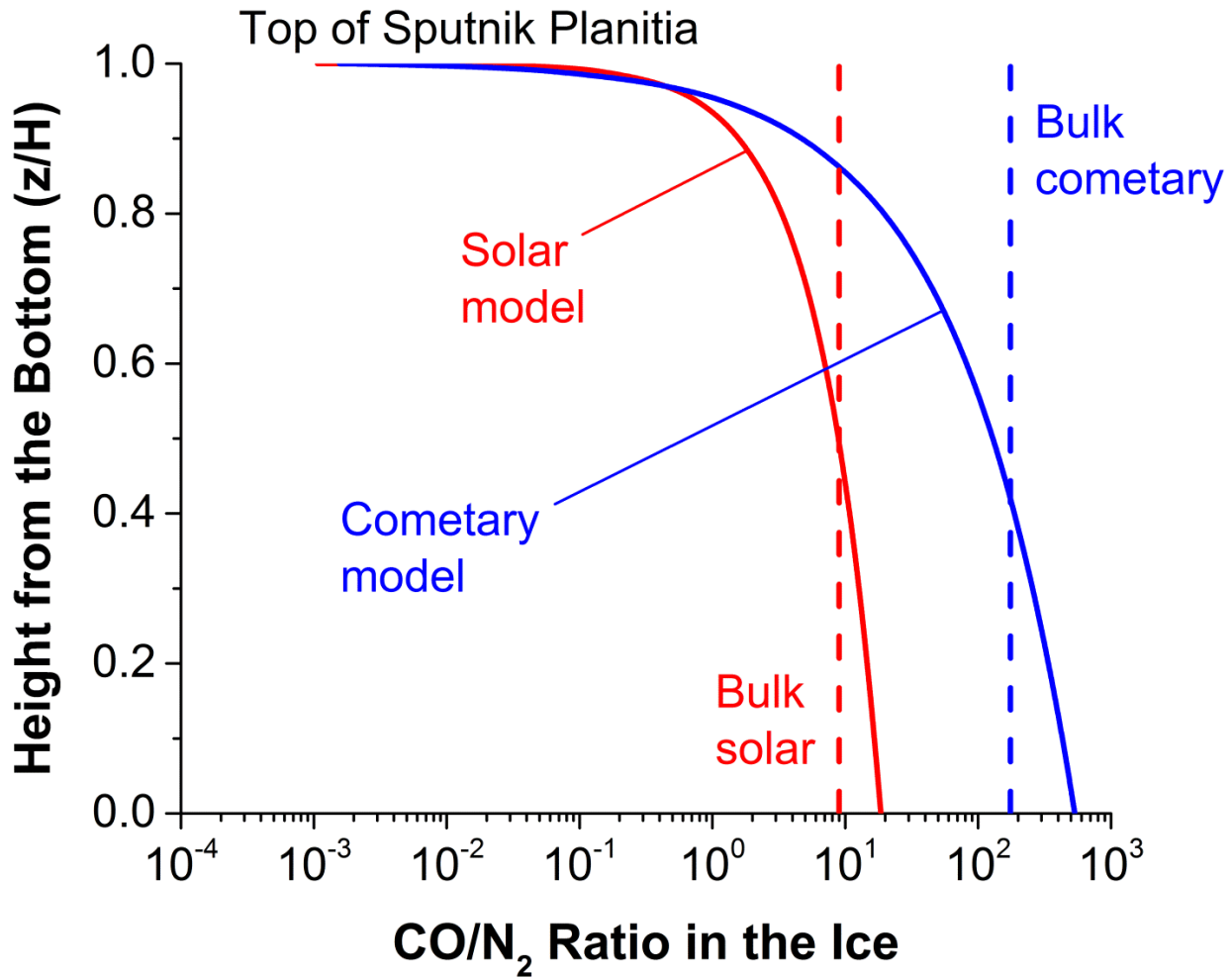

**Figure 5.** Model profiles of the ratio of $CO/N_2$ throughout the Sputnik Planitia ice sheet. These profiles were parameterized using Equation 7 to recover the atmospheric $CO/N_2$ ratio starting from a nominal cometary (blue) or solar (red) value.

Can observations of $CH_4$ be used to determine if volatiles are stratified as required by the preceding hypothesis? The spectroscopic model of Protopapa et al. (2017) indicates that the central portion of Sputnik Planitia contains about equal quantities of $N_2$- and $CH_4$-rich ices. Because $CH_4$ ice has a vapor pressure several orders of magnitude lower than that of $N_2$ on Pluto (Fray & Schmitt, 2009), it is tempting to regard the relatively high abundance of $CH_4$ as evidence against appreciable volatile layering. However, $CH_4$ could be unique in being enriched at the surface, despite its low volatility. This could be the case because the $N_2$-$CH_4$ system has a miscibility gap, which leads to the formation of $N_2$- and $CH_4$-rich phases on Pluto (Trafton, 2015). Because $CH_4$ ice (~0.5 g/cm$^3$) is considerably less dense than $N_2$ ice (~1 g/cm$^3$; Moore et al., 2016), the $CH_4$-rich phase may float through the low-viscosity $N_2$-rich phase to the surface. This possibility complicates the ability to relate the surface abundance of $CH_4$ to the bulk composition.

If CO is stratified in Sputnik Planitia, the profile of the $CO/N_2$ ratio would be controlled by volatility rather than density as CO ice is only ~1.5% denser than $N_2$ ice (Moore et al., 2016).

Owing to the buoyant behavior of $CH_4$-rich ice, a relatively high abundance of $CH_4$ at the surface does not preclude enrichment at depth for other species that are less volatile (or heavier) than $N_2$. Understanding the plausibility of intermediate layering of volatiles in Sputnik Planitia requires coupled chemical-dynamical modeling. It may be possible to test the idea of volatile layering if regions of upwelling from greater depths can be identified and correlated to higher surface abundances of less volatile species, such as CO.

*4.2. Destroying CO via aqueous chemistry*

A different way to obtain a low $CO/N_2$ ratio from an initial high $CO/N_2$ ratio is by chemically destroying CO while preserving $N_2$. This can happen if CO comes in contact with liquid water in the interior of an icy world (Shock & McKinnon, 1993). Laboratory rate data show that the destruction process is relatively rapid (Seewald et al., 2006), even at low temperatures if the aqueous solution has an alkaline pH (Trump & Miller, 1973). Using the latter kinetic data, Neveu et al. (2015) estimated that it may only take ~1-10 Myr to destroy 90% of the CO initially dissolved in liquid water. However, the kinetic data are applicable to far from equilibrium conditions, which leads to the more general question of whether thermodynamics allows the extreme degree of CO destruction that is required to reconcile the cometary (~175; see Section 3.2) and solar (~9; see Section 3.3) model ratios of $CO/N_2$ with that inferred at Pluto's surface (~$4 \times 10^{-3}$; see Section 4.1). Because the goal in this section is to search for a solution to the missing CO problem without invoking CO burial, it is assumed here that glacial ices in Sputnik Planitia are completely mixed by convection (McKinnon et al., 2016; Trowbridge et al., 2016), such that the equilibrium $CO/N_2$ ratio at the surface applies throughout the ice sheet. Assuming no production or preferential loss of $N_2$, the implication is that the present abundance of CO ice is consistent with a fraction of ~$2 \times 10^{-5}$ or ~$4 \times 10^{-4}$ of the initial cometary or solar model ratio, respectively.

Carbon monoxide reversibly interconverts with formate and carbonate species in geochemical environments containing liquid water (McCollom & Seewald, 2003; Seewald et al., 2006; McDermott et al., 2015). The carbon atoms in accreted CO on Pluto would be redistributed among these species if the CO were able to react with liquid water (e.g., in a subsurface ocean; Hammond et al., 2016; Johnson et al., 2016; Keane et al., 2016; Nimmo et al., 2016). A mass balance for this redistribution is

$$[CO]_0 \approx [CO]+[HCOOH]+[HCOO^-]+[CO_2]+[HCO_3^-]+[CO_3^{-2}], \qquad (10)$$

where brackets indicate the molal (mol/kg $H_2O$) concentration of the enclosed species. To quantify how this process could affect the budget of CO, we assume chemical equilibrium within the above set of species. The distribution of species in this model is a metastable distribution (akin to but more conservative than that in Shock & McKinnon, 1993), because it represents a local minimum in the Gibbs energy landscape of the chemical system. The considered set of species is a small subset of all possible C-O-H species that factor into stable equilibrium. Therefore, the metastable equilibrium concentration of CO provides an upper limit for the actual concentration of CO, because this simple model neglects other compounds (e.g., $CH_4$, organic materials, carbonate minerals) that could serve as sinks of CO-derived carbon (Zolotov, 2012; Neveu et al., 2017).

A set of reactions that defines metastable equilibrium between CO, formate, and carbonate species is given in Table 2. The equilibrium constants for these reactions were computed using the online version of SUPCRT92[1] with the slop07 database of revised Helgeson-Kirkham-Flowers (HKF) equation of state parameters (Johnson et al., 1992; Windman et al., 2008). The standard states that determine the values of these equilibrium constants are pure liquid water, and 1 molal solutions referenced to infinite dilution. SUPCRT92 uses the revised HKF equations (Tanger & Helgeson, 1988) to compute the standard-state Gibbs energy of individual species in a reaction at a specified pressure and temperature. The change in standard Gibbs energy for the reaction is calculated from the product and reactant Gibbs energies. The equilibrium constant is then obtained from the latter value (see Anderson, 2005 for an overview of the thermodynamic framework).

**Table 2.** Reactions in a model of metastable speciation of primordial CO in aqueous solutions on Pluto. The logarithms of the equilibrium constants are provided for representative conditions at the putative ocean floor of Pluto (McKinnon et al., 2017a).

| Reaction | log $K$ (0 °C, 1900 bar) |
|---|---|
| 1. $CO(aq) + H_2O(l) = HCOOH(aq)$ | 3.472 |
| 2. $CO(aq) + H_2O(l) = HCOO^-(aq) + H^+(aq)$ | -0.149 |
| 3. $CO(aq) + H_2O(l) = CO_2(aq) + H_2(aq)$ | 2.223 |
| 4. $CO(aq) + 2H_2O(l) = HCO_3^-(aq) + H^+(aq) + H_2(aq)$ | -3.836 |
| 5. $CO(aq) + 2H_2O(l) = CO_3^{-2}(aq) + 2H^+(aq) + H_2(aq)$ | -13.568 |

According to the law of mass action, the equilibrium constant at constant temperature and pressure can be expressed as the ratio of product/reactant activities ($a$), each of which is raised to its stoichiometric coefficient in the reaction. For example,

$$K_{R5} = \frac{a_{CO_3^{-2}} a_{H^+}^2 a_{H_2}}{a_{CO} a_{H_2O}^2} \approx \frac{[CO_3^{-2}] a_{H^+}^2 a_{H_2}}{[CO]}, \tag{11}$$

where we have introduced the approximation that the water is nearly pure, and the aqueous species exhibit ideal behavior. We have kept the activities of $H^+$ and $H_2$ in these forms because they are the fundamental compositional variables in the present problem, and further simplification is not needed. Expressions analogous to Equation 11 can be written for each of the reactions in Table 2. Those equations can then be rearranged to express the equilibrium concentration of the carbon-bearing product in terms of the other quantities. Continuing with the previous example, we have

$$[CO_3^{-2}] \approx \frac{K_{R5}[CO]}{a_{H^+}^2 a_{H_2}}. \tag{12}$$

---

[1] The online interface of SUPCRT92 can be accessed at geopig.asu.edu/tools.

This equation and its counterparts for the other carbon-bearing products can be substituted into Equation 10 to derive an expression for the fraction of primordial CO that remains in the aqueous solution at metastable equilibrium

$$\frac{[CO]}{[CO]_0} \approx \left(1 + K_{R1} + \frac{K_{R2}}{a_{H^+}} + \frac{K_{R3}}{a_{H_2}} + \frac{K_{R4}}{a_{H^+} a_{H_2}} + \frac{K_{R5}}{a_{H^+}^2 a_{H_2}}\right)^{-1}. \qquad (13)$$

    We find that thermodynamics strongly favors the destruction of CO under aqueous conditions that might have prevailed inside Pluto (Figure 6). This is consistent with the results of Shock & McKinnon (1993), who focused on the effects of temperature and oxidation state. Here, we focus on the effects of pH and oxidation state at the approximate melting temperature of water ice. We consider such a low temperature case to be conservative because forming cold (liquid) water demands less heating than making hot water, although hydrothermal evolutionary models for Pluto should be investigated in the future. An examination of Figure 6 reveals that the destruction of CO should be most pronounced under $H_2$-poor conditions at higher pH. This behavior can be rationalized by constructing an activity or predominance diagram (Figure 7), using the equilibrium constants in Table 2. At low activities of $H_2$ and high pH, the dominant species is the carbonate ion. It can also be seen that formate becomes the dominant species at elevated activities of $H_2$ and intermediate pH (Figure 7), which explains why the destruction of CO becomes independent of the $H_2$ activity in this pH regime (see Figure 6; there is no $H_2$ in Reaction 2 in Table 2). For reference, $a_{H2}$ in ultramafic-hosted hydrothermal fluids on Earth is of order $10^{-2}$ (Reeves et al., 2014).

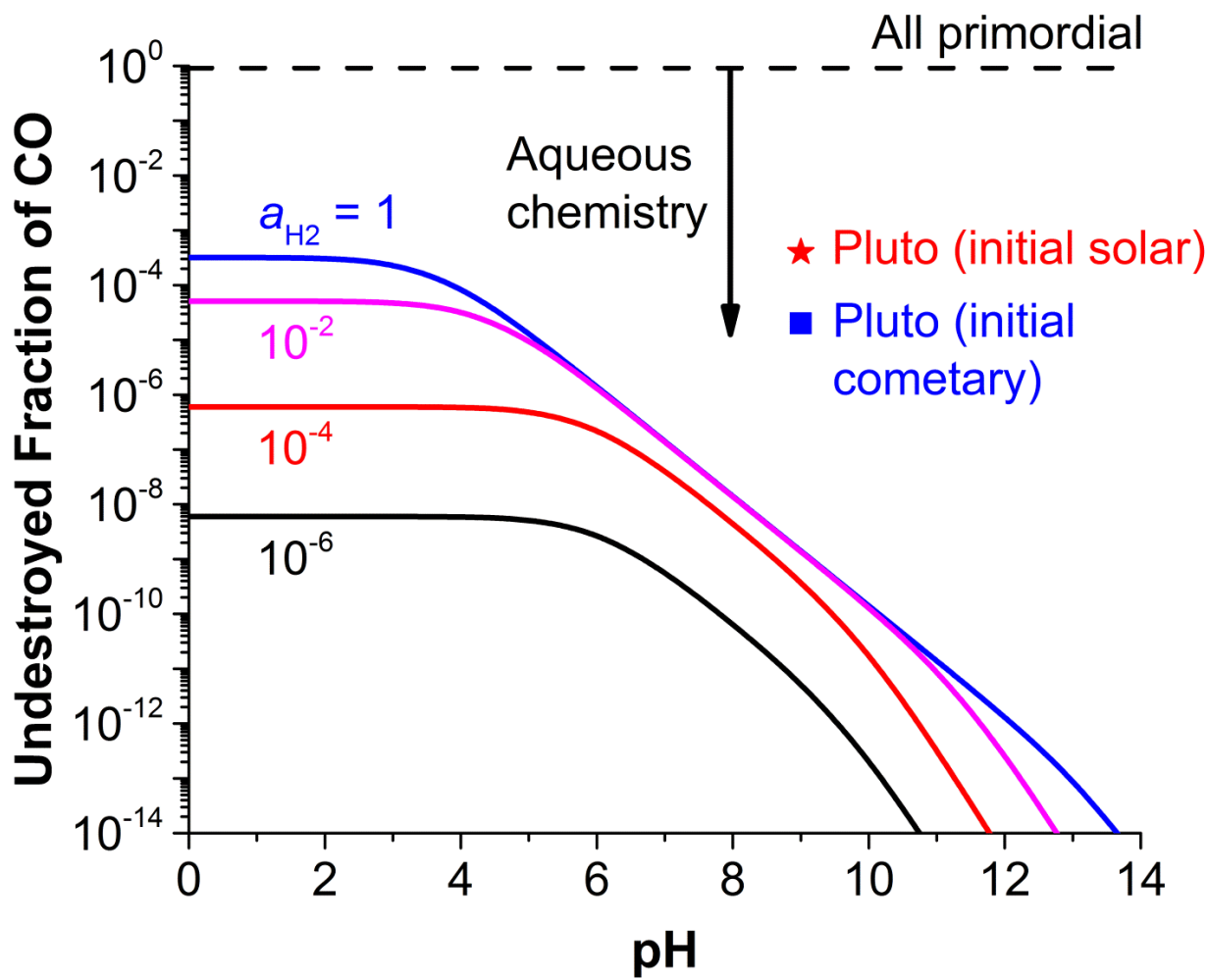

**Figure 6.** Fraction of initial CO that remains in aqueous solution at metastable equilibrium with formate and carbonate species at 0 °C and 1900 bar, as a function of the pH and $H_2$ activity (a measure of the oxidation state). The symbols labeled "Pluto" show how much CO destruction would need to occur for this model to reproduce the estimated surface ratio of $CO/N_2$ from an initial cometary or solar ratio (see Section 4.2). Here, the pH is assumed to be similar to a representative value for Enceladus' ocean (Postberg et al., 2009), though geochemical evaluations for Pluto have yet to be made.

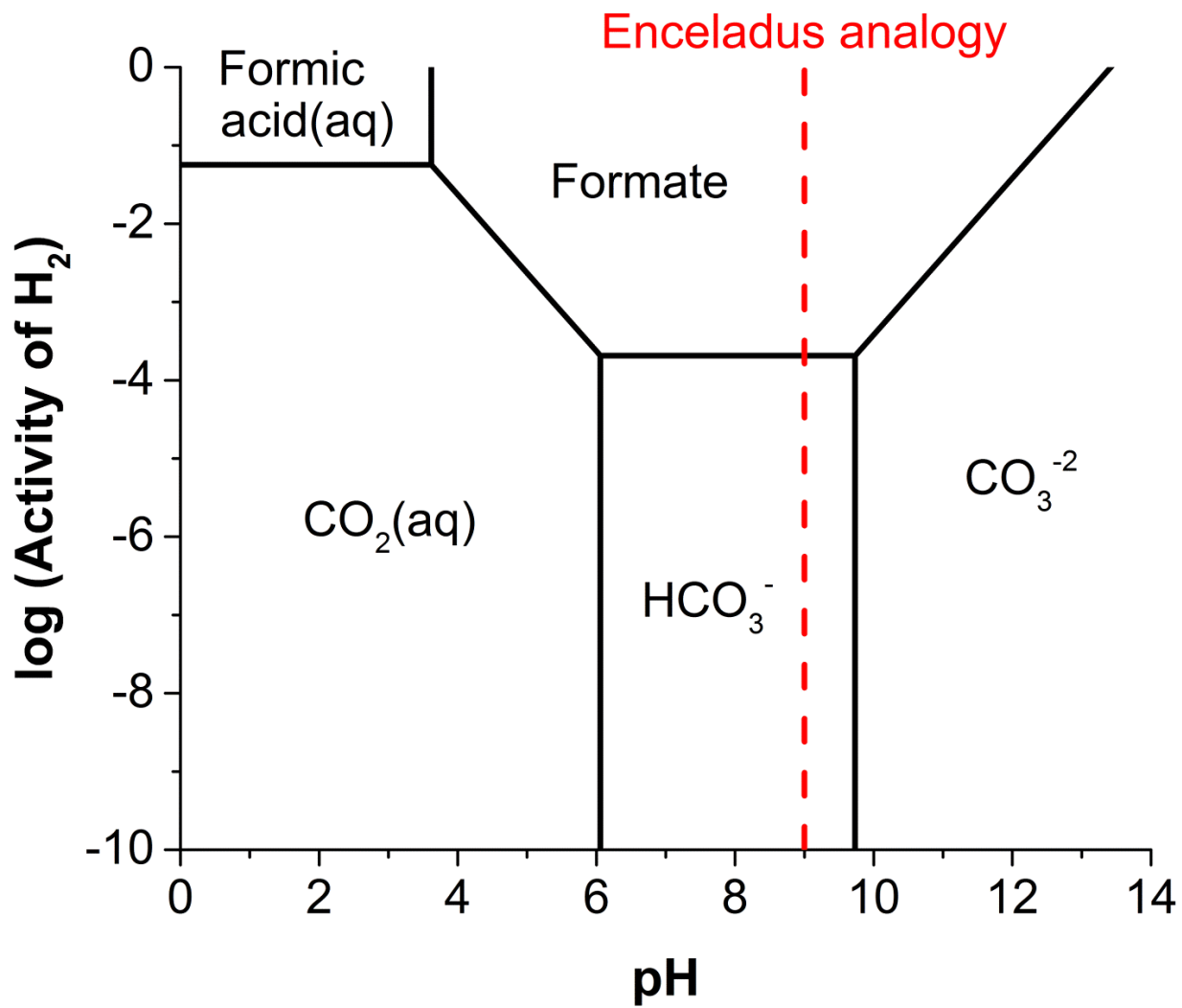

**Figure 7.** Regions of predominance between formate and carbonate species as a function of pH and $H_2$ activity (= molality in an ideal solution) at 0 °C and 1900 bar. The boundaries indicate where adjacent species have equal activities (~molalities). In its predominance region, the labeled species is the most abundant carbon compound in the metastable system. Carbon monoxide is never the most stable form of carbon anywhere in this diagram. The dashed red line provides an estimate of the pH if it is similar to a representative value for Enceladus' ocean (Postberg et al., 2009).

The metastable equilibrium model suggests that it is "easy" to decrease the $CO/N_2$ ratio by a large amount. But, can it quantitatively account for the data from Pluto? We have identified three scenarios that may provide consistency. First, the present CO abundance could be similar to the equilibrium abundance if the CO was derived from a relatively acidic fluid (Figure 6). This fluid does not necessarily need to be the suspected global ocean (Hammond et al., 2016; Johnson et al., 2016; Keane et al., 2016; Nimmo et al., 2016), and it may instead correspond to melts that formed during presumed water-rock differentiation on Pluto. However, maintaining a low pH is difficult for solutions in contact with silicate minerals as neutralization is favored (Zolotov, 2012; Neveu et al., 2017). If an alkaline pH is assumed, then the equilibrium would imply too much destruction of CO, even if the fluid was rich in $H_2$ (Figure 6).

In the second scenario, an apparent super-equilibrium abundance of CO on Pluto would be attributed to natural heterogeneity. One way for this to arise would be if a small fraction of accreted CO ($\sim 2\times 10^{-5}$ and $\sim 4\times 10^{-4}$ for the cometary and solar models, respectively) was never exposed to liquid water (Neveu et al., 2015). Another possibility is that some of the CO was protected in clathrate hydrates, even in the presence of liquid water. It is estimated that CO has a $\sim$50-fold preference in terms of concentration for a clathrate phase over being dissolved in aqueous solution (Davidson et al., 1987). This should lead to the preservation of more CO than implied by aqueous-only metastable equilibrium, for both thermodynamic and kinetic (because the rate of hydrolysis is proportional to the concentration of dissolved CO; Trump & Miller, 1973) reasons.

The third scenario assumes that effectively all of the accreted CO was destroyed under alkaline conditions, and Pluto's present CO is actually not primordial but was delivered by comets. Singer & Stern (2015) used the impact flux model of Bierhaus & Dones (2015), and estimated that $\sim 3\times 10^{17}$ kg of cometary materials have been delivered to Pluto over the past 4 billion years. If a typical comet is assumed to be composed of $\sim$50 wt. % $H_2O$ with a $CO/H_2O$ ratio of 1.4-8.8% (Dello Russo et al., 2016), then we might expect to find $\sim 10^{17}$-$10^{18}$ moles of comet-delivered CO at the surface of Pluto. For the estimated inventory of $N_2$ in Sputnik Planitia (Table 1) and a $CO/N_2$ ratio of $\sim 4\times 10^{-3}$ (see above), the corresponding amount of CO is $\sim(2-12)\times 10^{17}$ moles. The agreement between the amount predicted to be delivered by comets and the estimated amount on Pluto implies that a significant exogenic contribution of CO is plausible. An exception would be if a substantial quantity of CO was lost from Pluto, analogous to the LL model for $N_2$ in Section 2.2. While an external cometary source has the potential to account for the apparent CO inventory, the amount of N provided by this source (equivalent to $\sim 2\times 10^{17}$ moles of $N_2$; Singer & Stern, 2015) would not be sufficient to account for the corresponding inventory of $N_2$ on Pluto (see Table 1).

While the speciation calculations were consistent with a mass balance of CO-derived carbon, they did not include an oxygen balance which imposes an additional constraint. It can be deduced from Table 2 that 1-2 moles of $H_2O$ are required to destroy each mole of CO. This does not present a problem for the cometary model, since the typical range in the ratio of $CO/H_2O$ is 1.4-8.8% (Dello Russo et al., 2016). This model therefore provides a large excess of $H_2O$, which is consistent with water being a major component of Pluto's interior (see Section 3.1). In contrast, the endmember solar model is so CO-rich ($CO/H_2O \approx 2.6$; see Section 3.3) that $H_2O$ rather than CO is the limiting reactant in the destruction reactions (Table 2). This suggests that the solar model is incompatible with this mechanism of reconciling the $CO/N_2$ ratio, unless the excess CO was removed by accretional loss (Stevenson, 1993) prior to aqueous chemistry that fractionated $N_2$ from CO.

How can the aqueous destruction hypothesis be tested? One might expect to find evidence of the products of this chemistry such as $CO_2$, carbonate, or formate salts (Figure 7). Carbon dioxide has never been observed on Pluto's surface (Owen et al., 1993; Grundy et al., 2016a), although this could reflect an inefficiency in outgassing $CO_2$ rather than its absence in the interior. Carbon dioxide is many orders of magnitude less volatile than $N_2$, CO, and $CH_4$ (Fray & Schmitt, 2009). For carbonate or formate salts to be detected at the surface, salty water from a subsurface ocean would need to be emplaced on the surface (Neveu et al., 2015; Moore et al., 2016; Singer et al., 2016), and the salts not obscured by ice grains. Perhaps the simplest test of the present hypothesis that should be administered first is the prediction of a differentiated interior. This may be an inevitable outcome if there was extreme melting and aqueous processing of Pluto's primordial ices.

## 5. Concluding remarks

It was previously very difficult to study the chemical history of Pluto because of the paucity of relevant data. The *New Horizons* and *Rosetta* missions have changed the game by providing valuable new data, clearing paths to resolving this issue that is central to understanding the nature of Pluto. Here, we have journeyed down one of these paths by performing an in-depth investigation of whether Pluto's $N_2$ could be a primordial (accreted) species. We have emphasized the importance of mass balance, which relies upon a proper accounting of the atmospheric, escape, photochemical, and surficial reservoirs of $N_2$. A first attempt was made to quantify their inventories (Table 1) based on two different assumptions of past volatile loss (the Past Like Present, or PLP model; and the Large Loss, or LL model; see Section 2.2). Any hypothesis for the origin of nitrogen on Pluto must satisfy the top-down constraint that is imposed as much as possible by observational data.

We have also approached the mass balance from the bottom-up. This involves making an assumption for the $N_2$ content of the building blocks of Pluto. We considered two theoretical models: cometary and solar (see Sections 3.2 and 3.3, respectively). We found that the cometary model (informed by but not identical to comet 67P) exhibits an intriguing consistency with the PLP inventory, represented by $N_2$ ice in Sputnik Planitia (Figure 1). This defines a requirement of extensive outgassing of $N_2$ from the interior, and minimal escape of $N_2$ from Pluto's atmosphere. It is natural to compare the solar model to the LL inventory (Figure 2), as the former features a bountiful richness of $N_2$, albeit hypothetical in terms of existing solar system observations. In this case, consistency can be achieved for the opposite of the previous requirement, namely minimal outgassing or extensive escape (Figure 3). At the most basic level, we conclude that Pluto should have started with enough $N_2$. This provides support (although non-unique) for the primordial $N_2$ hypothesis, and invites independent scrutiny of which (if any) of the implied conditions are feasible. The implications of the two consistent cases for the evolution of Pluto are in stark contrast.

This brings us to the issue of missing CO, which must be confronted by any hypothesis that seeks to relate the composition of Pluto to those of more primitive bodies, such as comets. The conundrum entails explaining why relatively low $CO/N_2$ ratios are observed at Pluto (Owen et al., 1993; Lellouch et al., 2017). We recognize that one solution to the missing CO problem is that Pluto's $N_2$ is not primordial, but this demands an explanation of how primordial $N_2$ was either lost or not accreted by Pluto, and how secondary $N_2$ would be generated. Attracted by the simplicity of a primordial origin of $N_2$, we have chosen to explore the potential for co-accreted CO to go missing as a result of hypothesized processes on Pluto. First, we find that differences in vapor pressure could fractionate the $CO/N_2$ ratio between layers of ices at the surface (Figure 4). It is suggested that the ratio profile in Sputnik Planitia (e.g., Figure 5) may be controlled by a balance between differential volatility and convective overturning of glacial ices. Second, we have performed aqueous geochemical calculations showing the great thermodynamic instability of CO dissolved in cold (liquid) water, even for a restricted metastable equilibrium system (Figure 6). The destruction of CO to formate or carbonate species (Figure 7) would be strongly favored if Pluto has or had a subsurface ocean. This mechanism can be applied to the cometary model, but not to the solar model as the CO/H$_2$O ratio is too large in the latter. Hence, the cometary model seems preferable, with more options to reconcile the CO/N$_2$ ratio. We note that the burial and aqueous destruction hypotheses for missing CO are not necessarily mutually exclusive. A major implication of these processes is that the observed composition of Pluto cannot be completely primitive, even if its N$_2$ is indeed primordial. This resonates with the dynamic geology seen by *New Horizons* (Moore et al., 2016).

Speaking of a specific evolutionary scenario, we can envision the following: (i) Pluto started with cometary inventories of N$_2$ and CO; (ii) subsurface aqueous chemistry led to the destruction of CO; (iii) N$_2$ was subsequently outgassed efficiently; (iv) no significant loss of N$_2$ has occurred at the surface, and it accumulated in Sputnik Planitia; and (v) comets have delivered a small resupply of CO that mixes with N$_2$. This scenario is by no means the only possibility, but it is consistent with the present results.

This study of a primordial origin of Pluto's N$_2$ has led to an appreciation of many subsequent questions that must be addressed to develop a compelling model:

(1) Are other candidate sources of nitrogen (i.e., NH$_3$, organic materials; Lunine, 1993a) also consistent with the mass balance constraint in the context of present knowledge of conditions and processes on and within Pluto?
(2) What is the history of outgassing, atmospheric escape, and photochemistry (Mandt et al., 2016)? How variable have the rates of these processes been through time? What are the detailed mechanisms (e.g., of cryovolcanism), and how well can we constrain the governing physical and chemical properties?
(3) How much N$_2$ might be inside Pluto, and how might it be distributed among possible subsurface reservoirs (e.g., clathrate crust, liquid water ocean, rocky core)?
(4) Should we expect a giant impact to decrease Pluto's bulk N$_2$/H$_2$O ratio compared with primordial values (e.g., McKinnon, 1989)? Or, do the similar densities of Pluto and Charon (Nimmo et al., 2017) imply that volatiles had not yet segregated effectively to the surface at the time of the suspected Charon-forming impact?
(5) Is the N$_2$ abundance at comet 67P representative of other comets (e.g., Mousis et al., 2016) and larger icy protoplanets? Did solar composition material (excluding H$_2$, He, and Ne) condense anywhere in the early outer solar system (Owen & Encrenaz, 2006)?
(6) What is the compositional structure of Sputnik Planitia in three dimensions (e.g., homogeneous vs. highly stratified)? What processes control it, and how do their timescales compare? How much N$_2$ ice is present elsewhere on Pluto's surface (Protopapa et al., 2017)?
(7) What is the role of liquid water (cold and hot) in the evolution of the composition of volatiles on Pluto? How would interactions with rocks (Gabasova et al., 2018) affect the survivability of other primordial species such as O$_2$, which has not been observed (Kammer et al., 2017)? Could a sulfate-rich ocean be produced? If an ocean exists, what are its geochemical properties, such as pH and oxidation state (e.g., Waite et al., 2017)?
(8) Could the exterior inventory of CO on Pluto be exogenically derived? Perhaps from comet impacts or from a steadier influx of a source of oxygen, followed by its photochemical incorporation into CO (e.g., Hörst et al., 2008)?
(9) What does the lack of detection of CO$_2$ on Pluto but its presence on Triton (Cruikshank et al., 1993) mean with regards to resolving the missing CO problem on both bodies? Is surficial $CO_2$ on such objects an indicator of a geochemical endowment or geophysical processes (e.g., differing outgassing efficiencies)?
(10) Are there common mechanisms that provide volatiles to large icy worlds such as Titan, Triton, Pluto, and Eris? Or, might the unique pre- and post-accretionary histories of these bodies (Stern et al., 2018) determine their observable volatile inventories?

Some of these questions are amenable to additional analyses of the data from *New Horizons*, in coordination with theoretical modeling and laboratory experimentation. Others will remain recalcitrant until new data are acquired.

Considerable further progress in deciphering the origin of nitrogen on Pluto can be made by measuring

(1) The ratio of $^{14}N/^{15}N$ in HCN in Pluto's atmosphere. Lellouch et al. (2017) concluded that $^{14}N/^{15}N$ >125 based on the lack of detection of $HC^{15}N$ by ALMA. This raises the question of whether this species can be detected (or a stricter constraint obtained) if additional submillimeter observations are made. Relating the isotopic ratio in HCN to that in $N_2$ requires an understanding of photodissociative fractionation in the atmosphere of Pluto (Mandt et al., 2017). Modeling of this process can be substantiated using the known isotopic fractionation between HCN and $N_2$ on Titan (Niemann et al., 2010; Molter et al., 2016) as a test case.
(2) The ratio of $^{14}N/^{15}N$ directly in $N_2$ on Pluto. An atmospheric measurement would be better than nothing (Jessup et al., 2013), but may be difficult to relate to the bulk isotopic ratio at the surface owing to an extended altitude range of diffusive fractionation stemming from a low homopause (Young et al., 2018). While undoubtedly cost-challenging, the scientific interpretation would clearly benefit from making measurements at the surface. A mass spectrometer could be included on a Sputnik Planitia lander to provide this critical information. The $^{14}N/^{15}N$ ratios for endmember $N_2$, $NH_3$, and organic sources of Pluto's nitrogen are ~440 (Marty et al., 2011), ~130 (Rousselot et al., 2014), and ~225 (Miller et al., 2017), respectively. However, Pluto's $N_2$ could show some departure from these values as a result of atmospheric escape and the photochemical removal of heavy N over time (Mandt et al., 2016).
(3) The ratio of $^{36}Ar/N_2$ at Pluto's surface. Primordial $N_2$ (Owen, 1982; Balsiger et al., 2015) would be accompanied by a substantial inventory of $^{36}Ar$ (and $^{38}Ar$, the other primordial isotope). In contrast, secondary sources of $N_2$ would be consistent with a low $^{36}Ar/N_2$ ratio, as observed at Titan (Niemann et al., 2010). The measurement needs to be made on surface ices because diffusive fractionation in the atmosphere can be expected to be severe for argon. We may have already "seen" argon ice, but not yet recognized it as its effects on infrared spectra are subtle (Tegler et al., 2010). Complicating any interpretation would be the potential to bury Ar (see Section 4.1), or trap it in other materials (e.g., Mousis et al., 2013).
(4) The D/H ratio in $CH_4$ ice on Pluto. There must be a self-consistent story for the origin of all volatiles on Pluto, and not just $N_2$. This is where $CH_4$ can provide some guidance. If Pluto's $CH_4$ is an accreted species, then a similar primordial origin for $N_2$ would be made more plausible. Alternatively, a hydrothermal (Glein et al., 2008) or impact (Owen, 2000) origin for $CH_4$ would suggest that nitrogen species can also be subjected to energetic chemistry, allowing the production of $N_2$. A key constraint on the origin of CH$_4$ is its D/H ratio. It is of interest that CH$_3$D may have already been detected using ground-based spectroscopy (Protopapa et al., 2008). In the future, several CH$_3$D bands accessible to the James Webb Space Telescope can be used to determine the D/H ratio (Grundy et al., 2011).

We look forward to the power these measurements will bring to enable a deeper understanding of Pluto's chemical history.

**Appendix. Extrapolating present loss rates into the past**

Loss of material from atmospheres in the outer solar system is generally energy-limited. The solar fluxes of XUV (extreme ultraviolet) and Lyman-alpha photons are key drivers of the rates of escape and photochemistry (here, we do not consider impact erosion, which is beyond the scope of simple modeling; e.g., Griffith & Zahnle, 1995). As the Sun ages, these fluxes decrease, so to first-order the rates of escape and photochemistry can be expected to decrease through time. We attempt to roughly estimate the amount ($n$) of N$_2$ lost over the history of Pluto using the following scaling relationship between the rate of loss ($L$) and the combined high-energy photon flux ($I$) of XUV (1-1200 Å) and Lyman-alpha (1216 Å)

$$n_{lost} = \int_{t_1}^{t_2} (I/I_{today}) L_{today} dt, \tag{A1}$$

where $t$ represents time since solar system formation. The time dependence of the flux ratio in Equation A1 can be parameterized as (Ribas et al., 2005)

$$I/I_{today} = a_{XUV} t^{b_{XUV}} + a_{Ly\alpha} t^{b_{Ly\alpha}}. \tag{A2}$$

For $t$ in Earth years, $a_{XUV} = 3.16 \times 10^{11}$, $b_{XUV} = -1.23$, $a_{Ly\alpha} = 5.26 \times 10^6$, and $b_{Ly\alpha} = -0.72$. After inserting Equation A2, Equation A1 can be integrated analytically giving

$$n_{lost} = \left[ \left( \frac{a_{XUV}}{b_{XUV}+1} \right) \left( t_2^{b_{XUV}+1} - t_1^{b_{XUV}+1} \right) + \left( \frac{a_{Ly\alpha}}{b_{Ly\alpha}+1} \right) \left( t_2^{b_{Ly\alpha}+1} - t_1^{b_{Ly\alpha}+1} \right) \right] L_{today}. \tag{A3}$$

We restrict this analysis to $t_1 = 50 \times 10^6$ yr and $t_2 = 4.56 \times 10^9$ yr, as the parameterization was developed for ~0.1-7 Gyr (Ribas et al., 2005). However, the considered time interval covers ~99% of Pluto's presumed history. From Equation A3, we obtain

$$n_{lost} \approx (2.2 \times 10^{10} \text{ yr}) L_{today}, \tag{A4}$$

where the present loss rate is expressed as a quantity per year. This result implies that the mean loss rate over the 4.51 Gyr interval is ~5 times larger than the present rate. This scaling relationship can be used to constrain the total amount of N$_2$ that could have been removed from Pluto's atmosphere by escape and photochemistry.

**Acknowledgments**

C.R.G. wishes to thank Steve Desch, Jason Hofgartner, Jonathan Lunine, Mike Malaska, and Yasu Sekine for inspiring discussions on nitrogen on Pluto; Randy Gladstone for information on atmospheric escape from Pluto; Adrienn Luspay-Kuti and Kathy Mandt for sharing insights into N isotope fractionation on Pluto and Titan; and Kelly Miller and Ben Teolis for thoughtful discussions on a variety of topics related to the origin and evolution of volatiles in the outer solar system. C.R.G. dedicates this paper to the memory of P.J.G., who was an unsung hero in his life. We owe a debt of gratitude to three anonymous reviewers who helped to find ways to improve the quality of this work. We are also grateful to *New Horizons* for transforming Pluto from a speck of light to an extraordinary world, and with this revelation, opening the door to cosmochemical investigations of its origin. This project was supported by NASA *Rosetta* funding (JPL subcontract 1296001).